\documentclass[preprint,12pt]{elsarticle}

\usepackage{amssymb}
\usepackage{amsmath}

\usepackage{siunitx}
\usepackage{multirow}
\usepackage{makecell}
\usepackage[hidelinks,colorlinks=true,linkcolor=blue,urlcolor=blue,citecolor=blue]{hyperref}
\usepackage{placeins}
\usepackage{subcaption}
\usepackage{xcolor}
\usepackage[commandnameprefix=always, markup=default,authormarkup=none, final]{changes}

\definechangesauthor[name={Ruiyang Zhang}, color=orange]{rzhang}

\usepackage{lineno}

\journal{Nuclear Instruments and Methods in Physics Research Section A: Accelerators, Spectrometers, Detectors and Associated Equipment}

\begin{document}

\begin{frontmatter}



\title{Characterization of CharTPix-180A, a strip-shaped monolithic sensor for the STCF inner tracker}


\author[label2]{Ruiyang Zhang} 
\author[label2]{Dongwei Xuan}
\author[label2]{Hao Han}
\author[label2]{Xinyu Bin}
\author[label2]{Lailin Xu\corref{corre-author}}
\author[label2]{Jiajun Qin}
\author[label2]{Lei Zhao}
\author[label3]{Le Xiao}
\author[label3]{Xiangming Sun}
\author[label2]{Jianbei Liu\corref{corre-author}}
\cortext[corre-author]{Corresponding authors: Lailin Xu (lailinxu@ustc.edu.cn), Jianbei Liu (liujianb@ustc.edu.cn)}

\affiliation[label2]{
            organization={Department of Modern Physics},
            addressline={University of Science and Technology of China}, 
            city={Hefei},
            postcode={230026}, 
            country={China}}
\affiliation[label3]{
            organization={PLAC, Key Laboratory of Quark and Lepton Physics (MOE)},
            addressline={Central China Normal University}, 
            city={Wuhan},
            postcode={430079}, 
            country={China}}

\begin{abstract}

The Super Tau-Charm Facility (STCF) is a future high-luminosity electron-positron collider operating in the center-of-mass energy range of 2 to 7~\si{GeV}. The monolithic active pixel sensor (MAPS) is one of the options being considered for the STCF inner tracking detector. This application requires a MAPS capable of providing simultaneous position, time, and charge measurements under a constrained power budget of approximately \SI{50}{mW/cm^2}. To address these requirements, a baseline sensor concept featuring a strip-shaped pixel geometry has been proposed to balance power consumption and detection performance. The first prototypes of this concept, referred to as CharTPix-180A, have been developed to evaluate the proposed design. Two prototypes with different design focuses were characterized in this study. An analog readout prototype, CharTPix-180A-c4, was used to evaluate the intrinsic sensor behavior, while a fully functional digital readout prototype, CharTPix-180A-c3, was used to investigate the overall sensor functionality on a small scale. Electrical, radioactive-source, and beam tests were performed to characterize the prototypes. The measured detection efficiency exceeded 99\%, the fake-hit rate was below \SI{1e-3}{/pixel/s}, and the spatial resolution was better than \SI{10}{\micro\meter}. A best time resolution of \SI{37}{ns} was achieved, and the power consumption was about \SI{50}{mW/cm^2}. These results support the feasibility of the strip-shaped MAPS concept for the STCF inner tracker and provide guidance for further design optimization.

\end{abstract}

\begin{graphicalabstract}
\end{graphicalabstract}

\begin{highlights}
\item Developed the CharTPix-180A MAPS prototypes for the STCF inner tracker.
\item Achieved over 99\% detection efficiency with a fake-hit rate below \SI{1e-3}{/pixel/s} in the beam test.
\item Demonstrated the feasibility of strip-shaped sensor for the STCF inner tracker application.
\end{highlights}


\end{frontmatter}



\section{Introduction}
\label{intro}

Over the past few decades, the rapid development of monolithic active pixel sensors (MAPS) has enabled their widespread adoption in high-energy physics experiments, particularly in tracking and vertex detector systems~\cite{star,its2}. A key feature of MAPS lies in the monolithic integration of the sensing element and readout circuitry within a CMOS process. This approach takes advantage of advanced semiconductor manufacturing techniques, such as wafer stitching, allowing the realization of detectors with high spatial resolution, low material budget, low noise, and large active areas~\cite{dpts,moss}. In recent years, next-generation MAPS have been developed to provide not only hit position information but also additional information such as timing and deposited charge~\cite{picoad,color_run}.

The proposed Super Tau-Charm Facility (STCF) is a high-luminosity $e^+e^-$ collider designed to operate in the center-of-mass energy range of 2–\SI{7}{GeV}~\cite{CDR}. It aims to achieve a peak luminosity above $0.5\times10^{35}$~\si{cm^{-2}s^{-1}} at \SI{4}{GeV}, providing opportunities for precision studies of hadron structure and strong interactions, as well as searches for exotic hadrons and physics beyond the Standard Model.

Given the attractive features and promising potential of MAPS, it has been selected as one of the detector technology options for the inner tracking system of the STCF detector~\cite{CDR}. The envisaged inner tracker consists of four layers of MAPS covering radii from \SI{36}{mm} to \SI{196}{mm}. To meet the physics requirements of STCF, the MAPS design needs to balance spatial resolution, timing performance, and power consumption. In particular, the relatively moderate spatial resolution requirement of the STCF inner tracker provides room to optimize the pixel geometry and readout architecture for lower power consumption and improved timing performance. The sensors are expected to achieve a hit time resolution better than \SI{20}{ns} in order to reduce event pile-up. Charge measurement is important for improving timing resolution through time-walk correction and for helping the reconstruction of low-momentum tracks. Meanwhile, the sensor power consumption should be kept at approximately \SI{50}{mW/cm^2} to avoid the need for massive active cooling and realize a low material budget of 0.3\%~$X_0$ per detector layer. The combination of these requirements motivate dedicated R\&D on sensor structures and readout architectures tailored to the STCF inner tracker.

A previous simulation study has provided guidance for the design of MAPS for the STCF inner tracker~\cite{stcf_maps_sim}. The study indicates that adopting a strip-shaped pixel geometry can significantly enhance charge-collection performance, while the resulting degradation in front-end noise performance due to the increased sensor capacitance can be controlled at an acceptable level through circuit optimization. To experimentally validate this concept, an initial set of prototype chips was designed and fabricated using a \SI{180}{nm} CMOS imaging process~\cite{chartpix_180}. This process supports quadruple-well structures on a \SI{25}{\micro\meter}-thick epitaxial layer with a resistivity of greater than \SI{1}{\kilo\ohm\cdot cm}. The prototype series, comprising four different chip designs (c1–c4), is hereafter referred to as CharTPix-180A.

Among the four chip designs, an analog readout prototype, CharTPix-180A-c4, and a fully functional digital readout prototype, CharTPix-180A-c3, were characterized in this work~\footnote{CharTPix-180A-c1 and CharTPix-180A-c2 are being characterized separately, and the results will be reported elsewhere.}. CharTPix-180A-c4 features various sensor geometries combined with a simplified in-pixel front-end, providing an effective platform to study the intrinsic charge collection and analog performance of the strip-shaped sensor. In contrast, CharTPix-180A-c3 represents a small-scale prototype of the final chip and integrates all essential functionalities.

The experimental methods and results are presented in this paper as follows: The characterization of CharTPix-180A-c4 is described in Section~\ref{chip4}, including measurements of sensor capacitance and responses to X-rays ($^{55}$Fe) and beta particles ($^{90}$Sr). The performance of CharTPix-180A-c3 is discussed in Section~\ref{chip3}, covering power consumption and time resolution measurements using beta particles. Finally, Section~\ref{testbeam} presents the results from a dedicated beam test of CharTPix-180A-c3, providing key performance metrics such as detection efficiency and spatial resolution.

\section{Laboratory test of CharTPix-180A-c4}
\label{chip4}
\subsection{Overview of CharTPix-180A-c4}

The CharTPix-180A-c4 has an overall sensor area of $1.2\times 1.4$~\si{mm^2} and comprises three pixel matrices of at least $5\times5$ pixels each with distinct geometrical designs. 

\begin{figure}[htbp]
\centering
    \begin{subfigure}{0.4\textwidth}
        \centering
        \includegraphics[width=\linewidth]{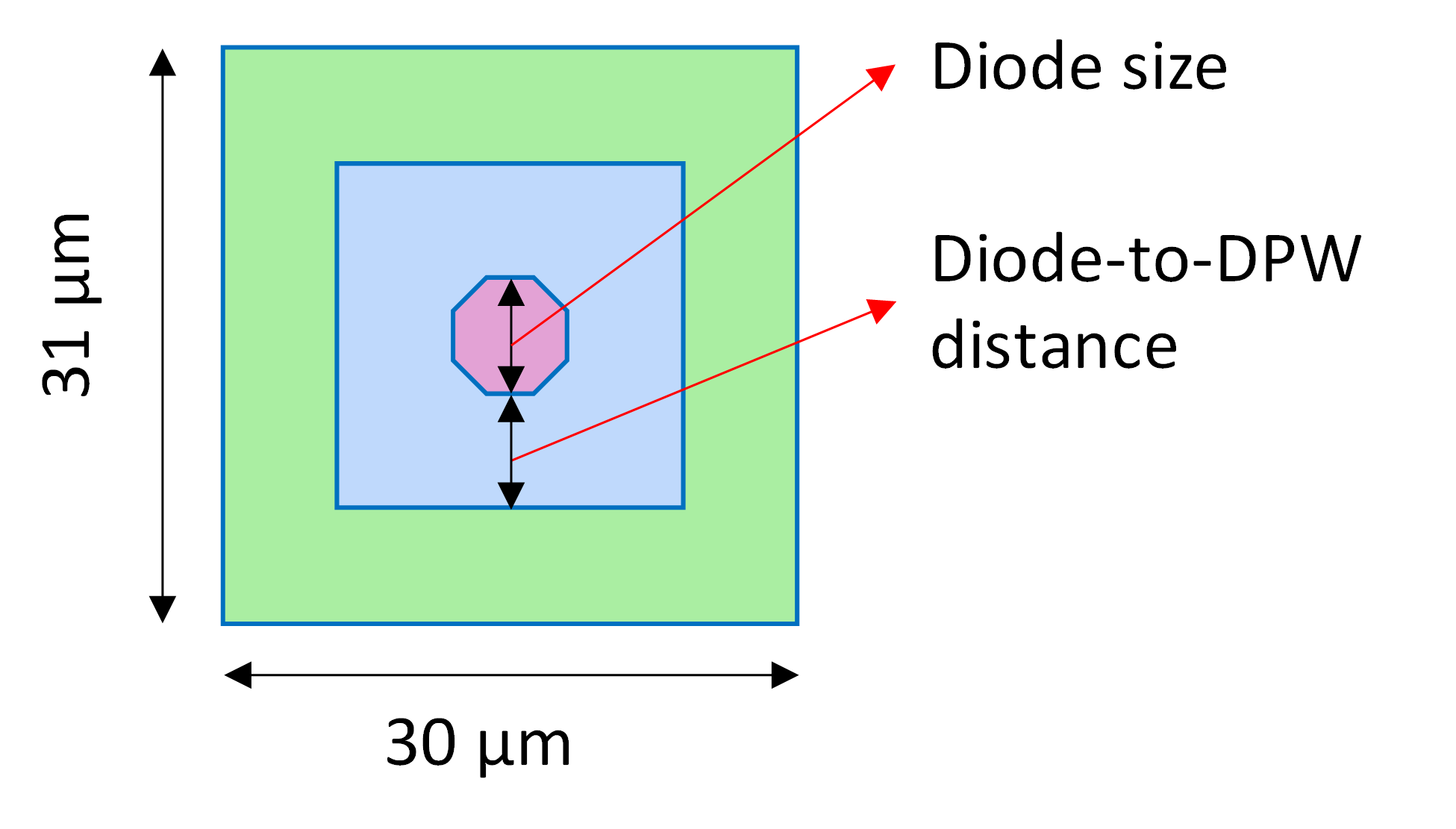}
        \caption{standard}
        \label{sensor_1}
    \end{subfigure}
    \hfill
    \begin{subfigure}{0.4\textwidth}
        \centering
        \includegraphics[width=\linewidth]{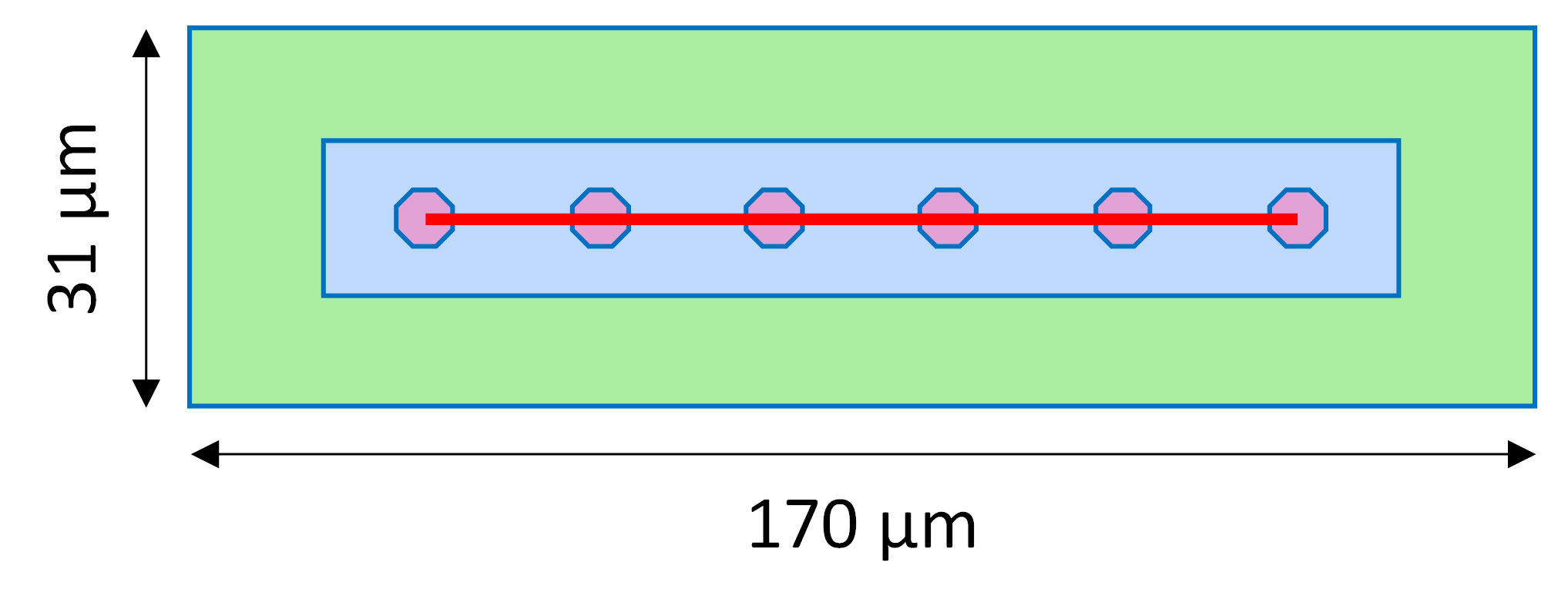}
        \caption{metal-connect}
        \label{sensor_3}
    \end{subfigure}
    \hfill
    \begin{subfigure}{0.4\textwidth}
        \centering
        \includegraphics[width=\linewidth]{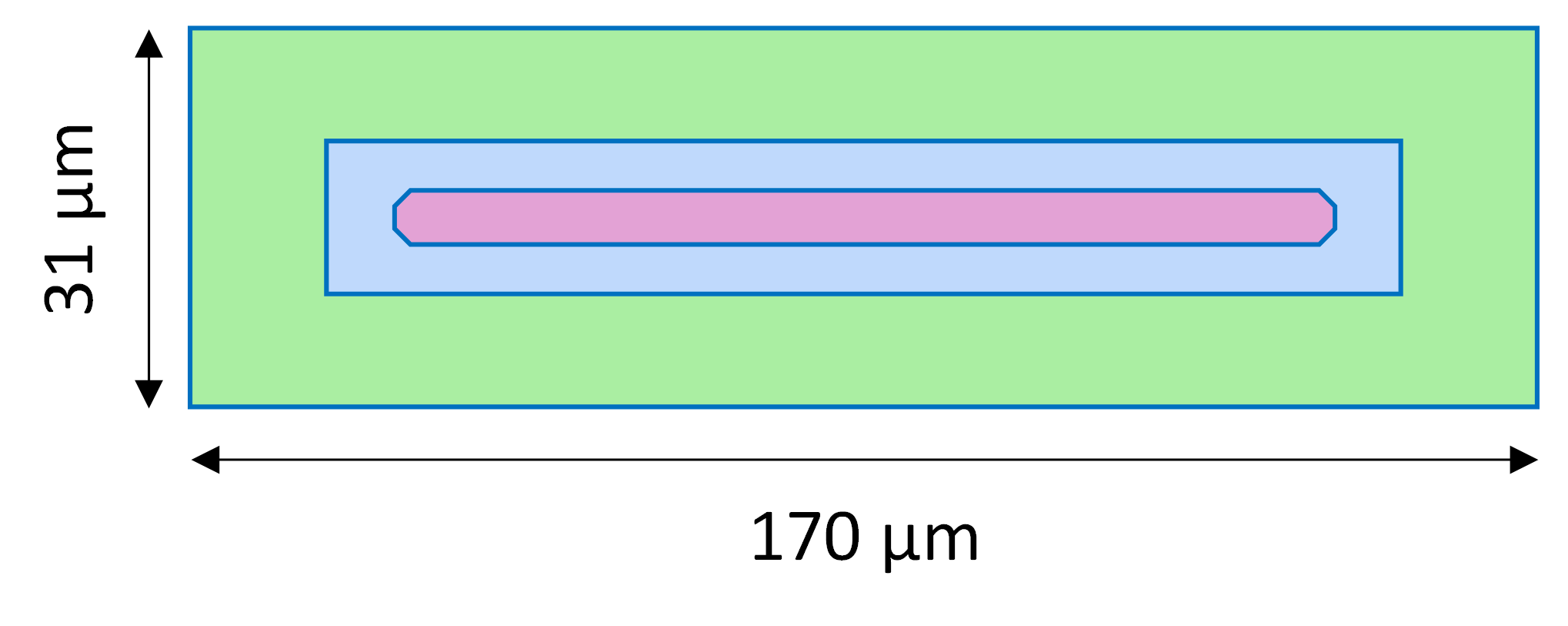}
        \caption{active-connect}
        \label{sensor_2}
    \end{subfigure}
\caption{Schematic layout of the three different pixel geometry variants in CharTPix-180A-c4: (a) standard, (b) metal-connect, (c) active-connect (not to scale). }\label{sensor_geometry}
\end{figure}

The first geometry variant (cf. Fig.~\ref{sensor_1}) is a conventional small-pitch sensor with a pixel size of $\SI{31}{\micro\meter} \times \SI{30}{\micro\meter}$. Each pixel contains a sensing diode with a lateral size of \SI{2}{\micro\meter}, separated by \SI{2}{\micro\meter} from the deep p-well (DPW) boundary. The other two sensor variants adopt an elongated pixel geometry with a pitch of $\SI{170}{\micro\meter} \times \SI{31}{\micro\meter}$, differing in their diode implementation. In the so-called ``metal-connect'' design (cf. Fig.~\ref{sensor_3}), six individual diodes within each pixel are connected by metal lines to form a single effective electrode. In contrast, the ``active-connect'' design (cf. Fig.~\ref{sensor_2}) employs a single elongated n-well diode spanning the pixel area. The detailed geometrical parameters of the three geometry variants are summarized in Table~\ref{tab1}. Minor differences with respect to the geometries used in the previous simulation study~\cite{stcf_maps_sim} arise from layout constraints in the final design.

In addition, a $5\times5$ null matrix without sensing diodes is implemented. These pixels retain the same front-end readout circuitry as the active pixels but are not connected to any charge-collecting electrode. This structure serves as a reference for separating the intrinsic sensor capacitance from parasitic contributions~\cite{soi}.

The DPW region of the entire chip is connected to an external substrate voltage $V_{sub}$ to provide reverse bias to the sensing diodes.

\begin{table}[htbp]
\centering
\caption{The main parameters for three pixel matrices.\label{tab1}}
\begin{tabular}{llllll}
\hline
Pixel type & Matrix size & Pixel size/\si{\micro\meter^2} & Diode size/\si{\micro\meter^2} & \makecell{Diode to DPW\\distance/\si{\micro\meter}} & \makecell{Diodes \\per pixel} \\
\hline
standard & $10\times10$ & $31\times30$ & $2\times2$ & 2 & 1 \\
active-connect & $5\times5$ & $170\times31$ & $140.375\times2$ & 2.52 & 1 \\
metal-connect & $5\times5$ & $170\times31$ & $2\times2$ & 2 & 6 \\
\hline
\end{tabular}
\end{table}

\begin{figure}[htbp]
\centering
\includegraphics[width=1.0\textwidth]{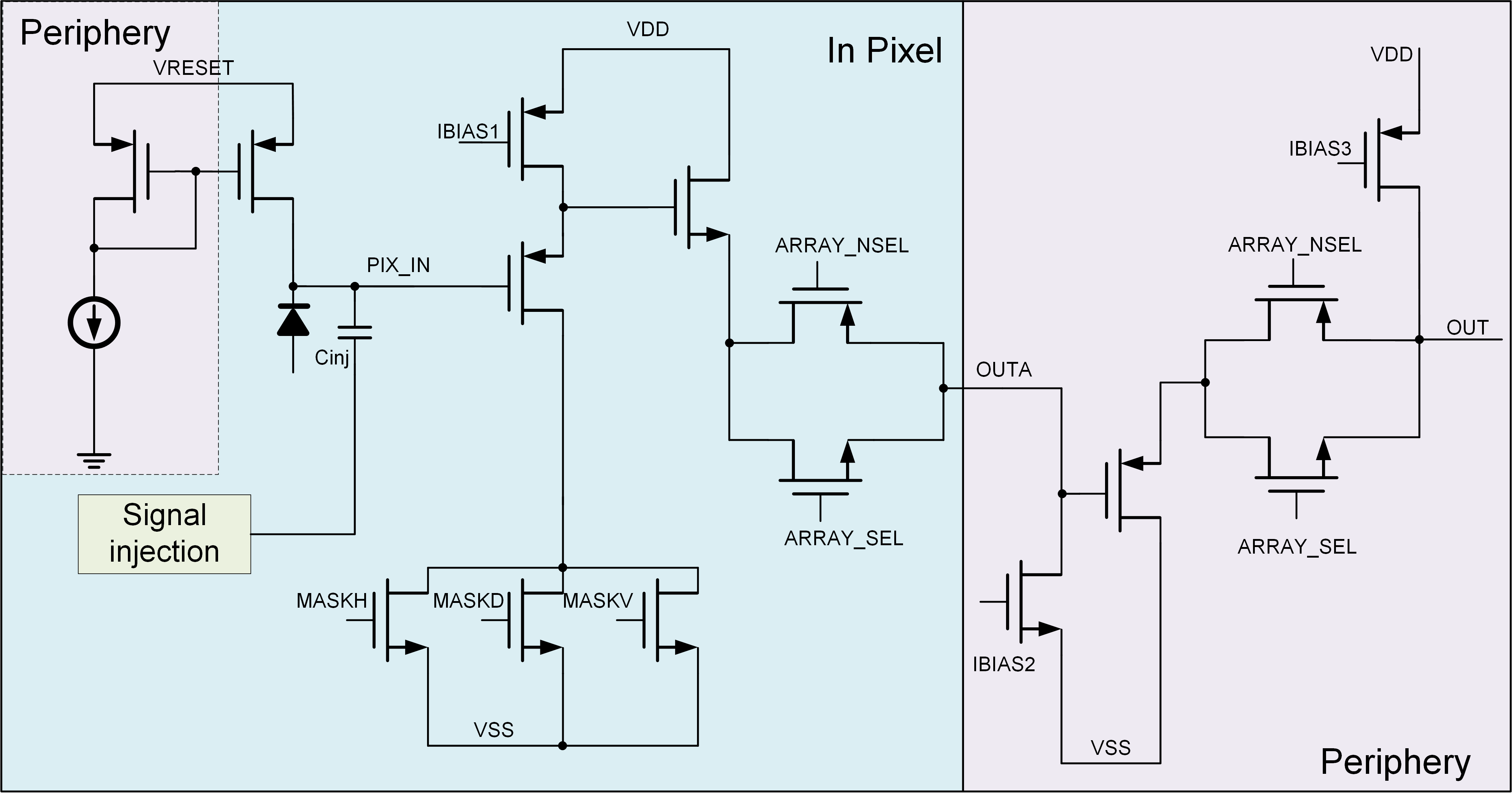}
\caption{Schematic diagram of the readout circuit for CharTPix-180A-c4.}\label{3T_front_end}
\end{figure}

Each pixel is equipped with a multi-stage source-follower-based readout circuit and an integrated charge injection circuit at the input node. Charge injection is implemented through applying a calibrated voltage step across a MIM capacitor with a nominal capacitance of approximately \SI{500}{aF} and slight variations among different pixel types. The sensing diode is reset via a PMOS transistor controlled by the $V_{\mathrm{reset}}$ signal, allowing the reset level to be adjusted. During operation, the signal from one selected pixel is transferred through the pixel-level readout stage and subsequently buffered by the peripheral readout circuitry to the output pad. In the presented measurements, the output was connected to an oscilloscope for waveform acquisition and analysis. A schematic diagram of the complete readout circuit is shown in Fig.~\ref{3T_front_end}.

\subsection{Capacitance measurements}

The input capacitance at the sensor input node, $C_{input}$, can be considered as the sum of the intrinsic sensor capacitance $C_s$ and the non-sensor-related capacitance $C_{ns}$, where $C_{ns}$ includes the input capacitance of the front-end circuitry and other parasitic contributions. In a null pixel, where no sensing diode is connected to the input node, the measured input capacitance provides an estimation of $C_{ns}$. The intrinsic sensor capacitance $C_s$ can therefore be obtained by subtracting $C_{ns}$ from the measured $C_{input}$ of an effective pixel.

The charge injection method was employed to extract the input capacitance:
\begin{equation}
    C_{input} \approx \frac{C_{inj}\cdot\Delta V_{pulse}}{\Delta V_{in}} 
    = \frac{C_{inj}\cdot\Delta V_{pulse}}{\Delta V_{out}/g_{sf}},
\end{equation}
where $C_{inj}$ denotes the injection capacitance and $\Delta V_{pulse}$ is the applied voltage step at the injection node\footnote{This approximation is justified by the circuit design, in which $C_{inj}$ is much smaller than $C_{input}.$}. The input node's voltage change $\Delta V_{in}$ cannot be measured directly and is inferred from the output voltage change $\Delta V_{out}$ through the gain of the source follower, $g_{sf}$.

The nominal value of $C_{inj}$ was initially obtained from circuit design simulations and subsequently calibrated using the full-energy peak position of a $^{55}$Fe source (see Section~\ref{c4_fe55}). The calibrated injection capacitance agrees with the design value within 10\% for all three pixel types. The source-follower gain $g_{sf}$ was determined via a DC sweep measurement, in which the output voltage was recorded for several input voltages around the nominal operating point. A linear fit to this transfer curve provided the gain as the slope of the fitted line.

\begin{figure}[htbp]
\centering
\includegraphics[width=1.0\textwidth]{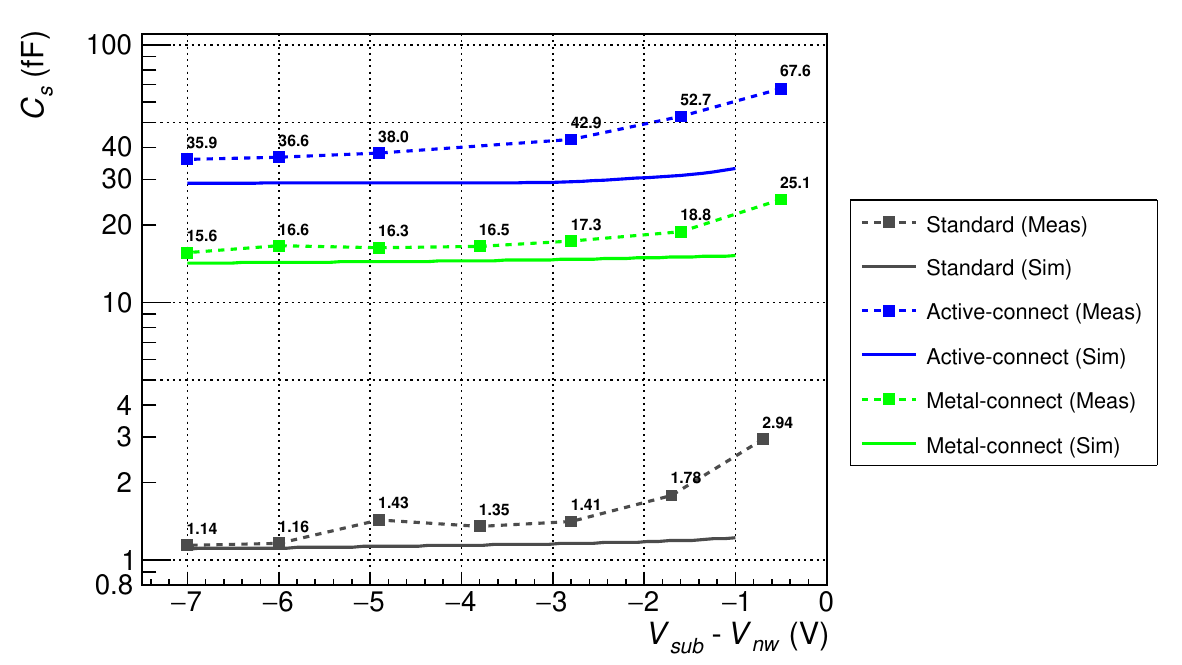}
\caption{Measured and TCAD-simulated sensor capacitance $C_{s}$ as a function of the reverse bias voltage $V_{sub}-V_{nw}$ for the three pixel types.}\label{cv_curve}
\end{figure}

After subtracting the parasitic contribution, the extracted sensor capacitance as a function of reverse bias voltage $V_{sub}-V_{nw}$ is shown in Fig.~\ref{cv_curve}. Among the three pixel variants, the active-connect sensor exhibits the largest capacitance and the strongest absolute variation with bias voltage, which is primarily attributed to its larger effective diode area.

For comparison, the corresponding TCAD-simulated capacitances are also presented in Fig.~\ref{cv_curve}. In the simulation, the doping profiles of the wells and epitaxial layer were obtained from material characterization of the fabricated chips. Nevertheless, a systematic deviation between simulation and measurement is observed, particularly at low substrate bias.

This discrepancy can be attributed to the different physical contributions captured by the two approaches. The TCAD AC simulation provides the small-signal junction capacitance of the sensing diode under fixed bias conditions. In contrast, the charge-injection method determines an effective capacitance from the injected charge and the resulting voltage response of the sensing node. The extracted capacitance therefore reflects not only the depletion capacitance of the sensing diode but also parasitic capacitance and bias-dependent capacitive coupling between the floating input node and the surrounding wells and readout structures. These additional contributions are expected to be more pronounced at low substrate bias and for elongated active-connect pixels, where the larger pixel geometry can enhance capacitive coupling to the surrounding structures. This results in a steeper apparent C–V dependence in the measurements. At sufficiently high reverse bias, the depletion region expands and the relative contribution of these additional coupling effects becomes smaller, leading to better agreement between the simulated and measured capacitance values.

\subsection{Energy spectrum with $^{55}$Fe X-ray}
\label{c4_fe55}
\begin{figure}[htbp]
\centering
\includegraphics[width=0.5\textwidth]{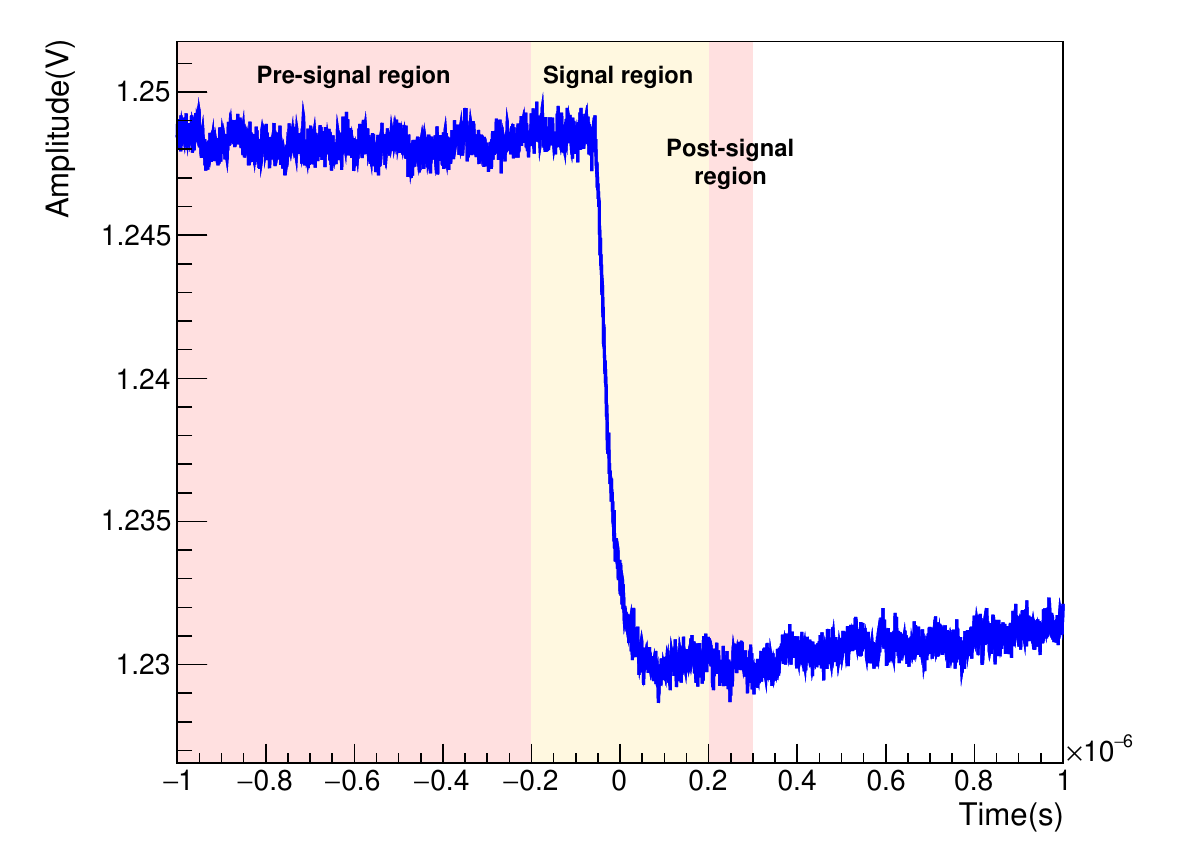}
\caption{A typical signal waveform from the standard pixel type, irradiated by a $^{55}$Fe source.}\label{waveform}
\end{figure}

The charge collection capability and energy resolution of the sensor were studied using irradiation with a $^{55}$Fe X-ray source. The source was placed on the front side of the sensor without any intervening material, in order to minimize photon absorption by non-sensitive layers. Since only one pixel can be read out during the measurement, a central pixel from each sensor matrix was selected and connected to an oscilloscope as a representative of the corresponding sensor geometry. A trigger threshold was set at $5\sigma$ above the baseline noise level to suppress noise hits. Fig.~\ref{waveform} shows a typical signal waveform recorded by the oscilloscope. Owing to the fast rising edge and comparatively slow signal decay, the waveform within the acquisition window can be divided into three regions: a pre-signal region preceding the rising edge, a signal rising window, and a post-signal region following the rising edge.

The pre-signal region was selected before the signal onset to determine the baseline level. The signal rising window was defined as the narrowest common interval that fully contained the rising edges of the tested signals. A subsequent \SI{100}{ns} interval was used to determine the post-signal waveform level, thereby limiting the influence of the subsequent slow signal decay.  The signal amplitude was then determined from the difference between the average waveform levels of the pre-signal and post-signal regions.

\begin{figure}[htbp]
\centering
    \begin{subfigure}{0.5\textwidth}
        \centering
        \includegraphics[width=\linewidth]{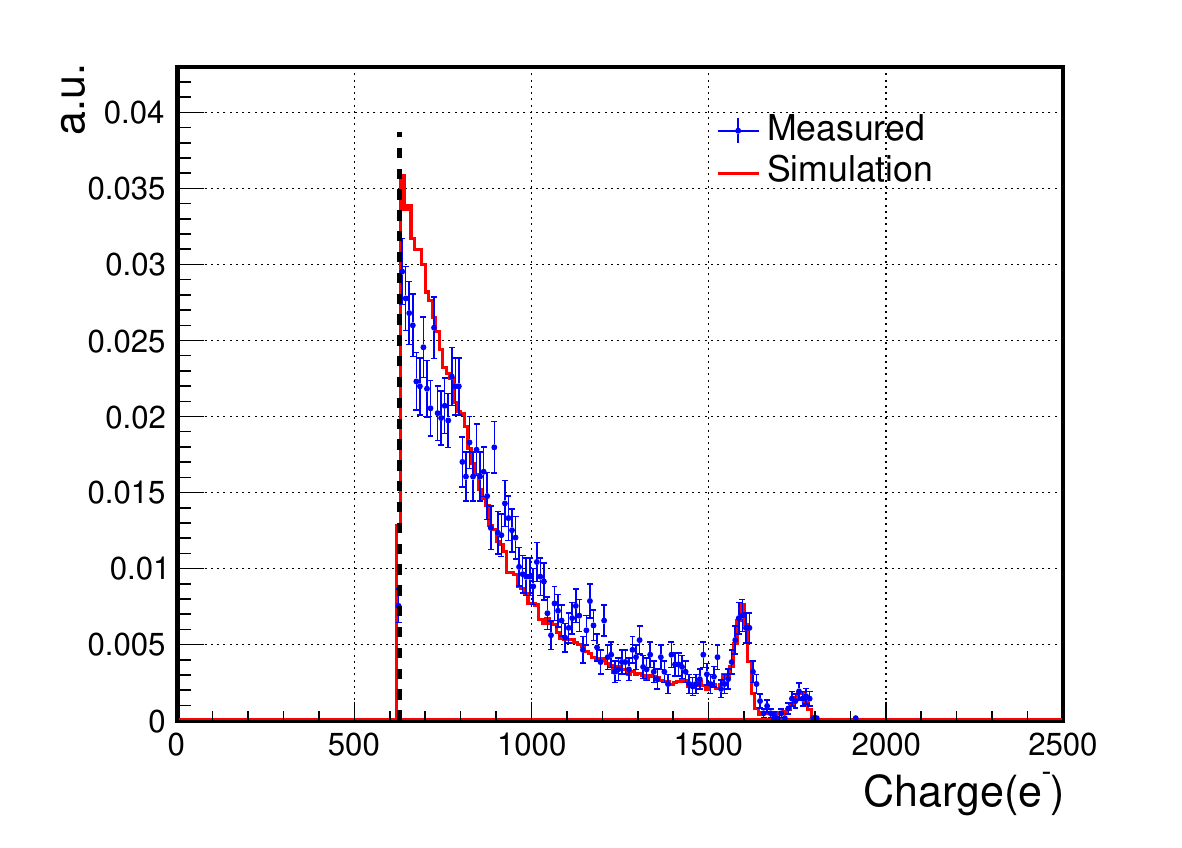}
        \caption{standard}
        \label{spectrum_1}
    \end{subfigure}
    \hfill
    \begin{subfigure}{0.5\textwidth}
        \centering
        \includegraphics[width=\linewidth]{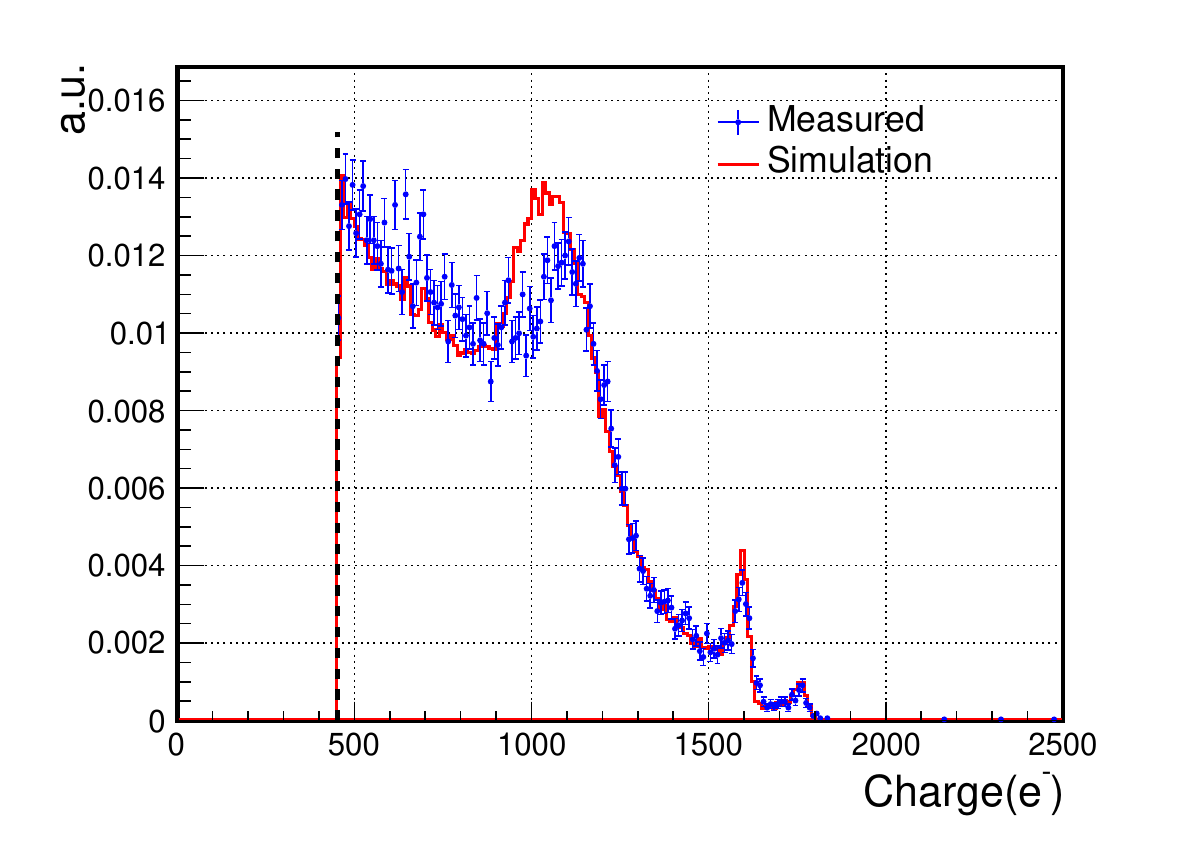}
        \caption{metal-connect}
        \label{spectrum_3}
    \end{subfigure}
    \hfill
    \begin{subfigure}{0.5\textwidth}
        \centering
        \includegraphics[width=\linewidth]{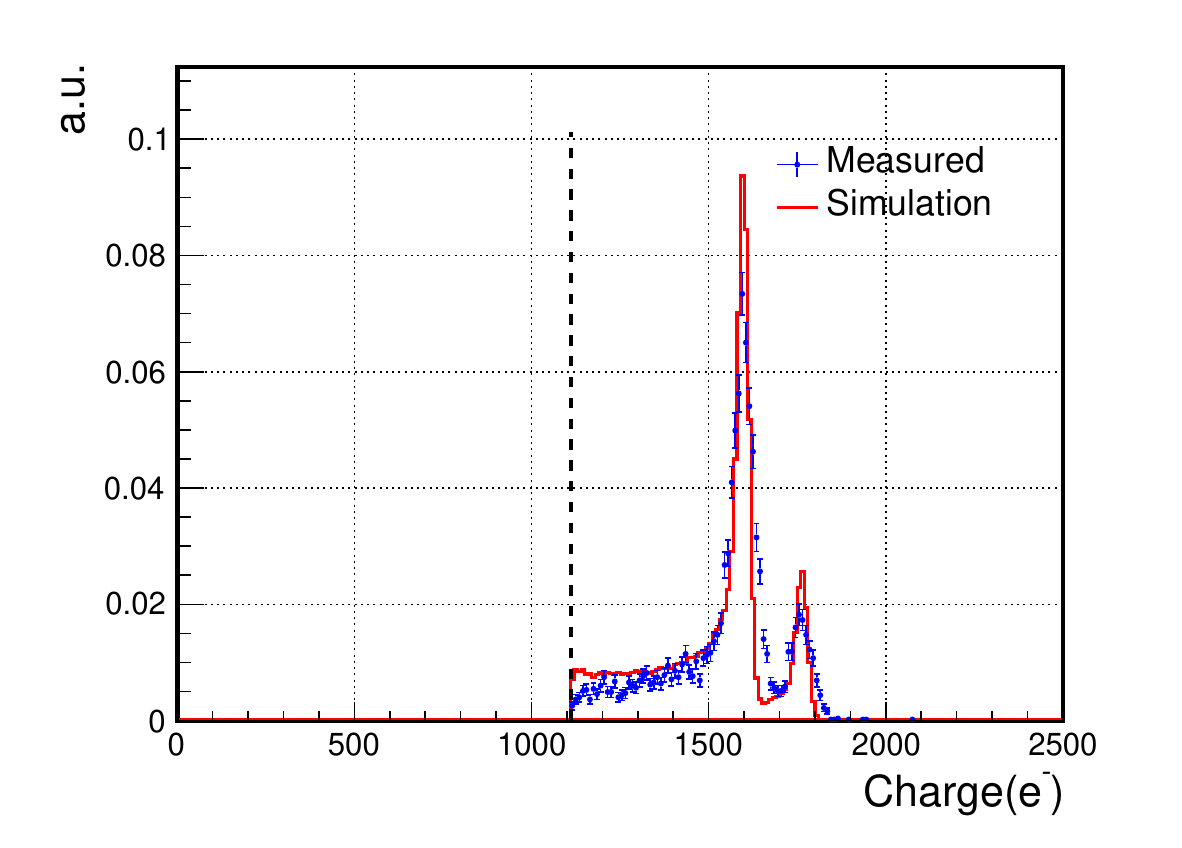}
        \caption{active-connect}
        \label{spectrum_2}
    \end{subfigure}
\caption{Comparison of the measured and simulated $^{55}$Fe energy spectra for a single pixel of three sensor variants: (a) standard sensor, (b) metal-connect sensor, and (c) active-connect sensor. The dashed black line indicates the applied threshold during data acquisition.}\label{spectrum_compare}
\end{figure}

The energy spectrum was obtained by accumulating the signal amplitudes of all triggered events. The measured spectra for the three sensor variants at a substrate bias voltage of \SI{-6}{V} are shown in Fig.~\ref{spectrum_compare}. The raw signal amplitudes were converted into collected charge by calibrating the full-energy peak position to the energy deposited by the $^{55}$Mn-$K_{\alpha}$ line in silicon, corresponding to \SI{1620}{e^-}. The oscilloscope trigger threshold is indicated as the dashed line. In all cases, the $^{55}$Mn-$K_{\alpha}$ and $^{55}$Mn-$K_{\beta}$ peaks are clearly resolved.

A qualitative comparison of the charge collection performance can be made from the relative prominence of the full-energy peak with respect to the low-amplitude region, which originates from partial charge collection due to charge sharing between neighboring pixels. Among the tested designs, the active-connect sensor exhibits the best charge collection performance, as indicated by the highest fraction of events in the full-energy peak. This behavior is consistent with its larger depletion region, which enhances charge collection efficiency.

To validate the simulation framework developed within the STCF offline software environment~\cite{oscar,stcf_maps_sim}, the simulated energy spectra were compared with the experimental data. In the measurement, the readout pixel is not necessarily the seed pixel (i.e., the pixel collecting the largest charge in a cluster) when multiple pixels exceed the threshold. This effect was taken into account in the simulation by randomly selecting the charge from one of the pixels above threshold. The simulated spectra show good agreement with the measured data over most of the amplitude range above threshold. A slightly broader full-energy peak is observed in the measurement for the active-connect sensor than in the simulation. This difference is partly attributed to the noise introduced by the analog readout circuit, which is not included in the simulation. The large input capacitance of the active-connect sensor further increases the impact of readout noise on the signal-to-noise ratio.

\subsection{Timing performance}
\label{timing_performance}


The timing performance of different sensor variants was studied by analyzing the signal rise time, defined as the time interval required for the signal amplitude to increase from 10\% to 90\% of its final value.

\begin{figure}[htbp]
\centering
    \begin{subfigure}{0.5\textwidth}
        \centering
        \includegraphics[width=\linewidth]{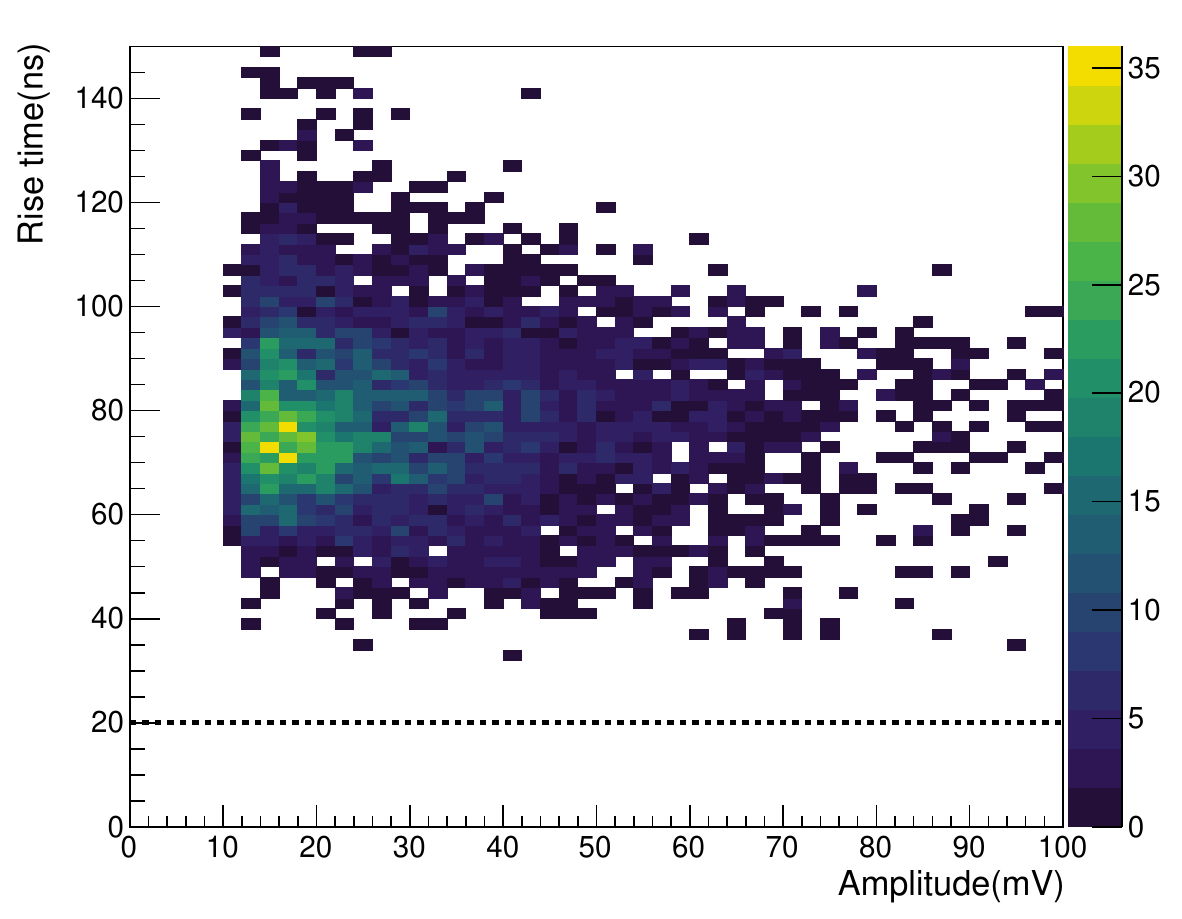}
        \caption{}
        \label{rise_time_1}
    \end{subfigure}
    \hfill
    \begin{subfigure}{0.5\textwidth}
        \centering
        \includegraphics[width=\linewidth]{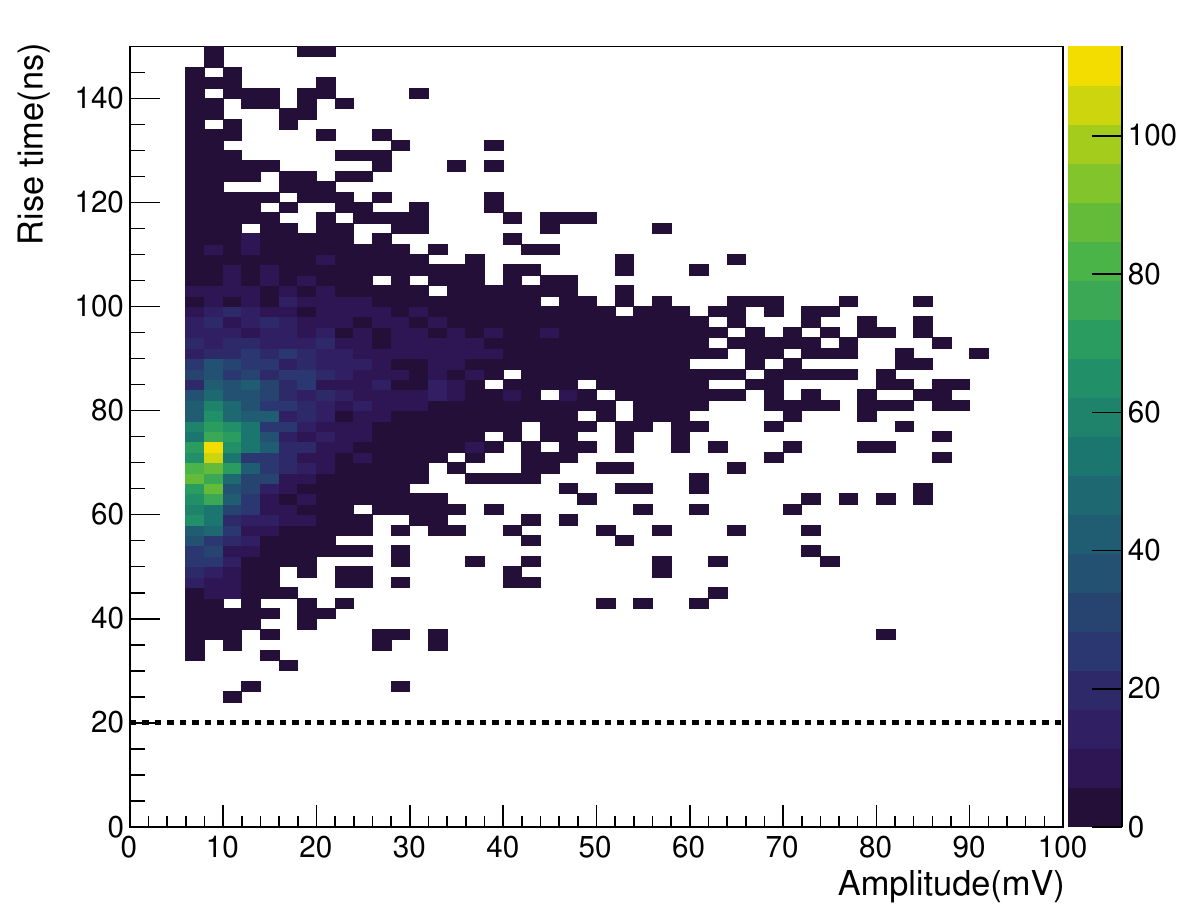}
        \caption{}
        \label{rise_time_2}
    \end{subfigure}
    \hfill
    \begin{subfigure}{0.5\textwidth}
        \centering
        \includegraphics[width=\linewidth]{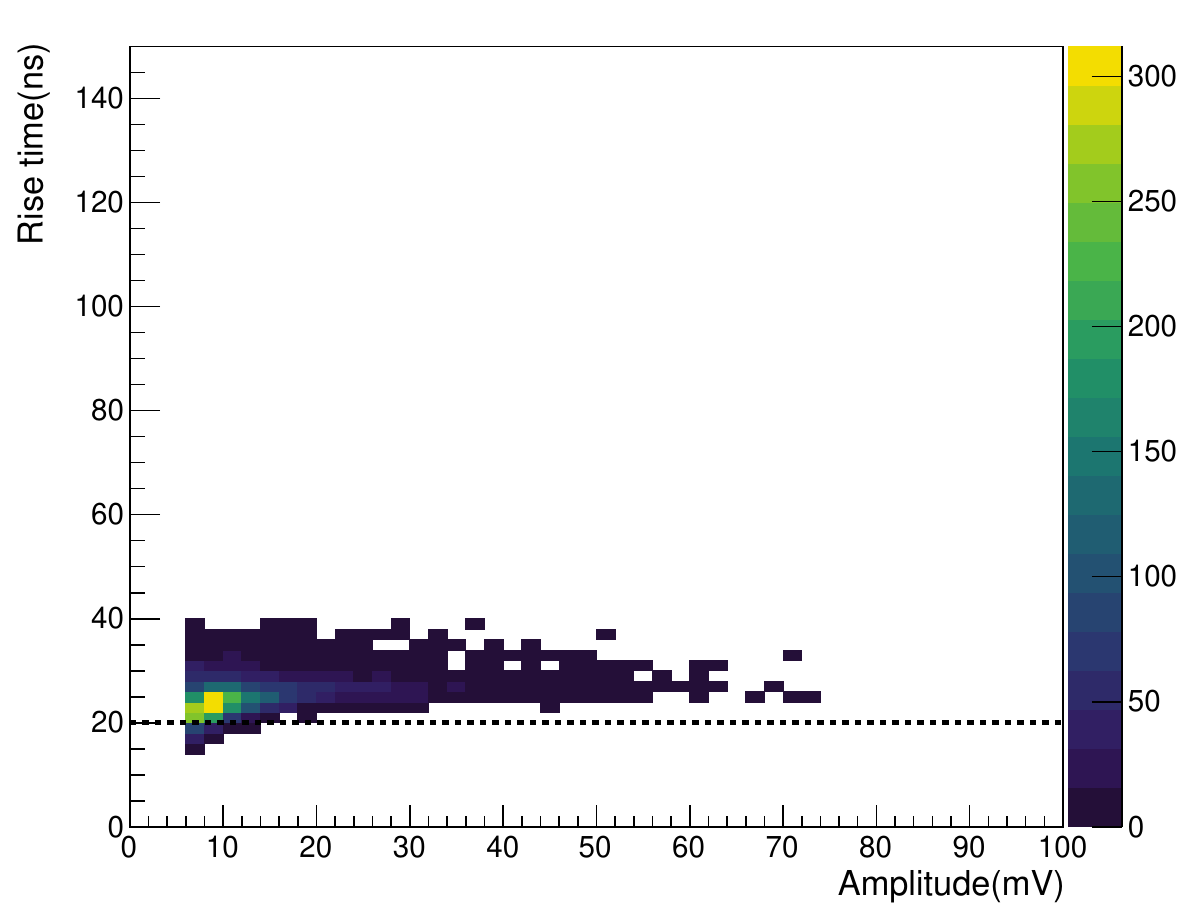}
        \caption{}
        \label{rise_time_3}
    \end{subfigure}
\caption{Signal rise time versus signal amplitude of $^{90}$Sr source for: (a) standard sensor, (b) metal-connect sensor and (c) active-connect sensor }\label{rise_time}
\end{figure}

Fig.~\ref{rise_time} shows the rise time of the three sensor variants under $^{90}$Sr $\beta$ source  irradiation as a function of the calibrated signal amplitude (in $e^-$). Owing to the limited bandwidth of the source-follower readout circuit, the minimum achievable rise time was estimated to be approximately \SI{20}{ns}. This limit was verified by both circuit simulations and charge injection measurements and is indicated by a dashed line in the plots. For the standard sensor and the metal-connect strip sensor (cf. Fig.~\ref{rise_time_1} and Fig.~\ref{rise_time_2}), the measured rise time was significantly longer than the circuit limit and showed a larger spread at certain signal amplitudes. This behavior was attributed to the slower charge collection process in these sensors, resulting from their relatively limited depletion regions. In contrast, the active-connect sensor (cf. Fig.~\ref{rise_time_3}) exhibited a rise time that was nearly independent of the signal amplitude and remained consistently limited by the readout circuit. This observation suggested that the intrinsic rise time of this sensor was well below \SI{20}{ns}, providing strong support for achieving the targeted timing performance required for the STCF inner tracker.

A rise-time measurement using an $^{55}$Fe X-ray source was also performed. Since conclusions consistent with those obtained from the $^{90}$Sr measurements were observed, the detailed results are not presented here. A more quantitative determination of the time resolution of the CharTPix-180A-c4 chip was not feasible within the scope of the current setup, as the requirement of single-pixel readout during the measurement limited the achievable data rate. The time-resolution performance of the CharTPix-180A-c3 chip was therefore studied separately and is discussed in Section~\ref{time_res_c3}.

\section{Laboratory test of CharTPix-180A-c3}
\label{chip3}

\subsection{Overview of CharTPix-180A-c3}
\label{CharTPix-c3-overview}

The CharTPix-180A-c3 chip is a fully functional MAPS prototype developed toward the realization of a low-power sensor with integrated time and charge measurement capabilities. The chip has an overall size of $2.8 \times 3.1$~\si{mm^2} and contains a pixel matrix of 15 columns by 60 rows. The matrix is divided into two distinct sectors implementing different strip-like sensor designs: the lower seven columns adopt the metal-connect scheme, while the upper eight columns employ the active-connect scheme. The diode size and diode-to-DPW distance are identical to those used in the CharTPix-180A-c4 prototype, while the pixel pitch is slightly modified to $\SI{169}{\micro\meter} \times \SI{31}{\micro\meter}$ (cf. Table~\ref{tab1}).

Each pixel integrates an open-loop analog front-end and a digital logic block capable of recording 8-bit leading-edge (LE) and trailing-edge (TE) timestamps. The front-end bias settings are shared among pixels within the same sector. Pixel hits are read out using a column-drain architecture, in which hits are prioritized locally within the matrix. To suppress spurious signals from hot pixels, three mask latches, corresponding to horizontal, vertical, and diagonal masking, are provided to selectively disable individual pixels. A detailed description of the chip architecture can be found in Ref.~\cite{chartpix_180}.

However, the current prototype chip exhibited an output signal oscillation effect. Further investigation revealed that this behavior originated from a strong parasitic coupling between the front-end output node and its input node. As a result, an additional charge of approximately \SI{1200}{e^-} was injected back into the input once the signal crossed the discrimination threshold, causing the discriminator output to be clipped immediately after the leading edge. This effect was consistently reproduced in full-size post-layout simulations, confirming its origin in the present circuit routing. It is therefore expected that this issue can be resolved in a future design iteration through layout optimization.

To ensure the correct operation of the basic characterization measurements with the current prototype, an off-chip reset logic was implemented to automatically reset the entire pixel matrix after a hit was detected. While this workaround allowed stable operation, it introduced two technical limitations in the chip characterization. First, the signal clipping prevented the application of a time-walk correction or TA calibration. Second, the global reset mechanism introduced a chip-level dead time after each pixel hit, during which subsequent hits did not generate a response. The reset window was minimized to approximately \SI{20}{\micro s}; however, it still affected the test results under high-count-rate conditions. A dedicated treatment in the data analysis was therefore applied to account for this effect and to enable a reliable performance evaluation (see Section~\ref{testbeam}).

\begin{figure}[htbp]
\centering
\includegraphics[width=0.8\textwidth]{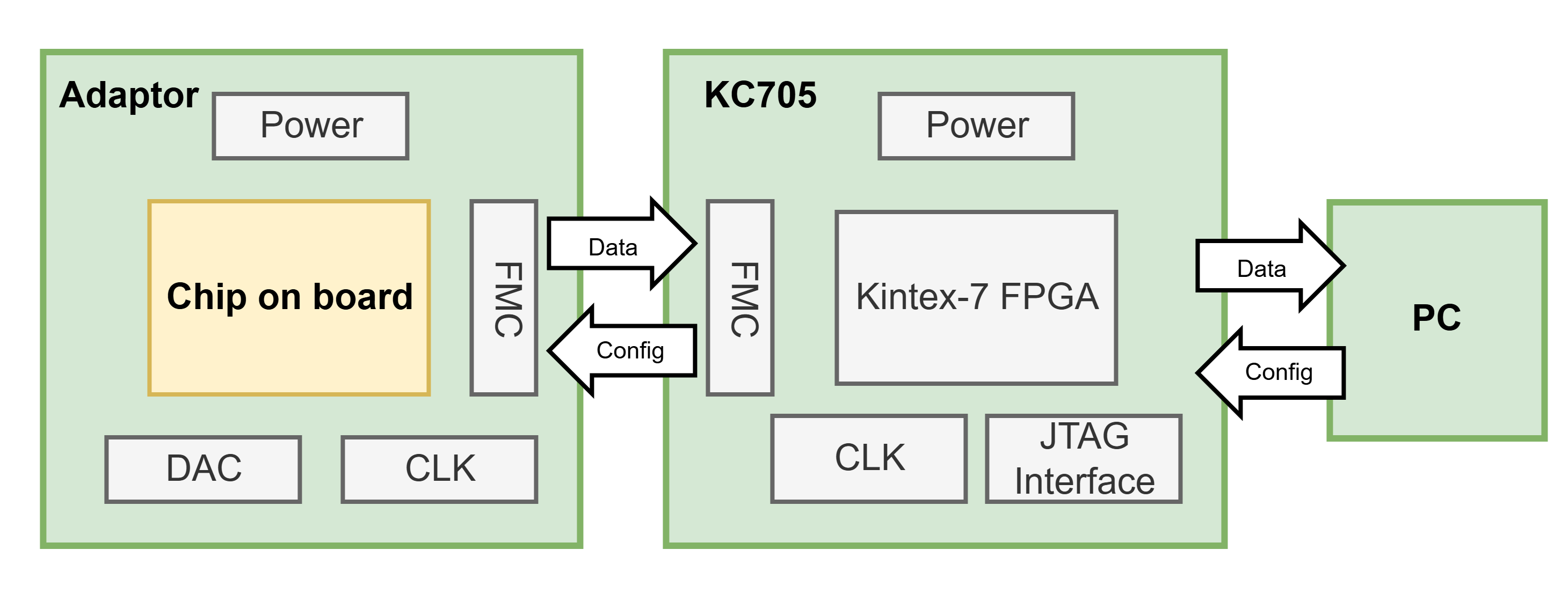}
\caption{A schematic view of the characterization setup used in laboratory tests.}\label{test_system}
\end{figure}

The chip characterization setup consisted of a chip-carrier board, an adapter board, and an FPGA board (cf. Fig.~\ref{test_system}). The chip-carrier board was specifically developed to host the device under test (DUT) and to provide the necessary passive components for chip powering and communication. The carrier board was mounted on a universal adapter board that supplied the off-chip bias voltages required by the chip. A dedicated FPGA firmware was implemented on a KC705 development board for slow control and data readout. The configuration, slow control, and data acquisition processes were managed through a software interface based on the IPbus framework~\cite{ipbus}.

\subsection{Power consumption}

To meet the stringent low-power requirement of the STCF inner tracker, power consumption measurements were performed to evaluate the effectiveness of the adopted power-optimization strategies~\cite{chartpix_180}. Since CharTPix-180A-c3 has a relatively small chip area, some power contributions cannot be directly extrapolated to the target full-size chip of $2\times2$~\si{cm^2}. In particular, the dynamic power associated with data readout and the power consumption of the peripheral circuitry depend strongly on the overall chip size and architecture. In contrast, the analog front-end power within the pixel matrix and the power associated with timestamp distribution can be normalized to the pixel area and extrapolated to a full-size chip based on the measured values.

For the analog front-end power measurement, the front-end bias current ($I_\mathrm{BIAS}$) was gradually increased from 0 to \SI{1500}{nA} by adjusting the on-chip DAC setting. Over this range, the total current consumed by the analog front-end circuits of the active-connect and metal-connect pixel matrices increased by \SI{0.505}{mA} and \SI{0.296}{mA}, respectively. After converting these current increments to power and normalizing by the corresponding pixel matrix area, the analog power consumptions were determined to be \SI{36.9}{mW/cm^2} and \SI{24.3}{mW/cm^2} for the active-connect and metal-connect structures, respectively.

Timestamp distribution power is another important contributor to power consumption that is targeted by the power-optimization strategy of CharTPix-180A-c3. With timestamp distribution enabled, the measured total digital current was \SI{4.31}{mA}, compared with \SI{3.85}{mA} when the distribution was disabled. The difference of \SI{0.46}{mA} corresponds to a timestamp-distribution power consumption of \SI{17.8}{mW/cm^2} after normalization to the pixel matrix area. This measured value is in good agreement with the corresponding simulation result.

The total power density was estimated by summing these two contributions, yielding \SI{54.7}{mW/cm^2} and \SI{42.1}{mW/cm^2} for the active-connect and metal-connect structures, respectively. The estimated power density of the metal-connect structure is below the STCF target of \SI{50}{mW/cm^2}, while that of the active-connect structure is slightly higher. These results demonstrate the effectiveness of the power-optimization strategy, particularly for the metal-connect structure. However, the estimates include only the two contributions that can be reliably extrapolated from the prototype and do not account for dynamic readout power or peripheral circuitry power. Further optimization and evaluation of these contributions are therefore needed to determine whether the final chip can meet the overall power budget.

\subsection{Time resolution}
\label{time_res_c3}

The time resolution of the two sensor types implemented in the CharTPix-180A-c3 chip was investigated using a $^{90}$Sr beta source. The source was placed above the MAPS chip under test, while an LGAD sensor was positioned beneath the sensor and served as the time reference detector. The LGAD sensor consists of $5\times5$ pads, each with a size of $1.3\times1.3$~\si{mm^2}, and could be independently amplified and read out~\cite{LGAD_sensor_design, LGAD_electronics_design}. In this study, however, the channels from all pads were combined through an OR logic to provide a sufficiently large sensitive area.

The LGAD signal was then routed to the FPGA board of the DUT, where its leading edge was sampled at a frequency 16 times higher than the MAPS system clock and used as the reference time. This configuration resulted in a time bin size of \SI{3.125}{ns}. Since the intrinsic time resolution of the LGAD had been determined to be better than \SI{100}{ps} under the corresponding bias condition, the overall uncertainty of the reference time was estimated to be approximately \SI{0.9}{ns} (\SI{3.125}{ns}$/\sqrt{12}$).

\begin{figure}[htbp]
\centering
\includegraphics[width=0.5\textwidth]{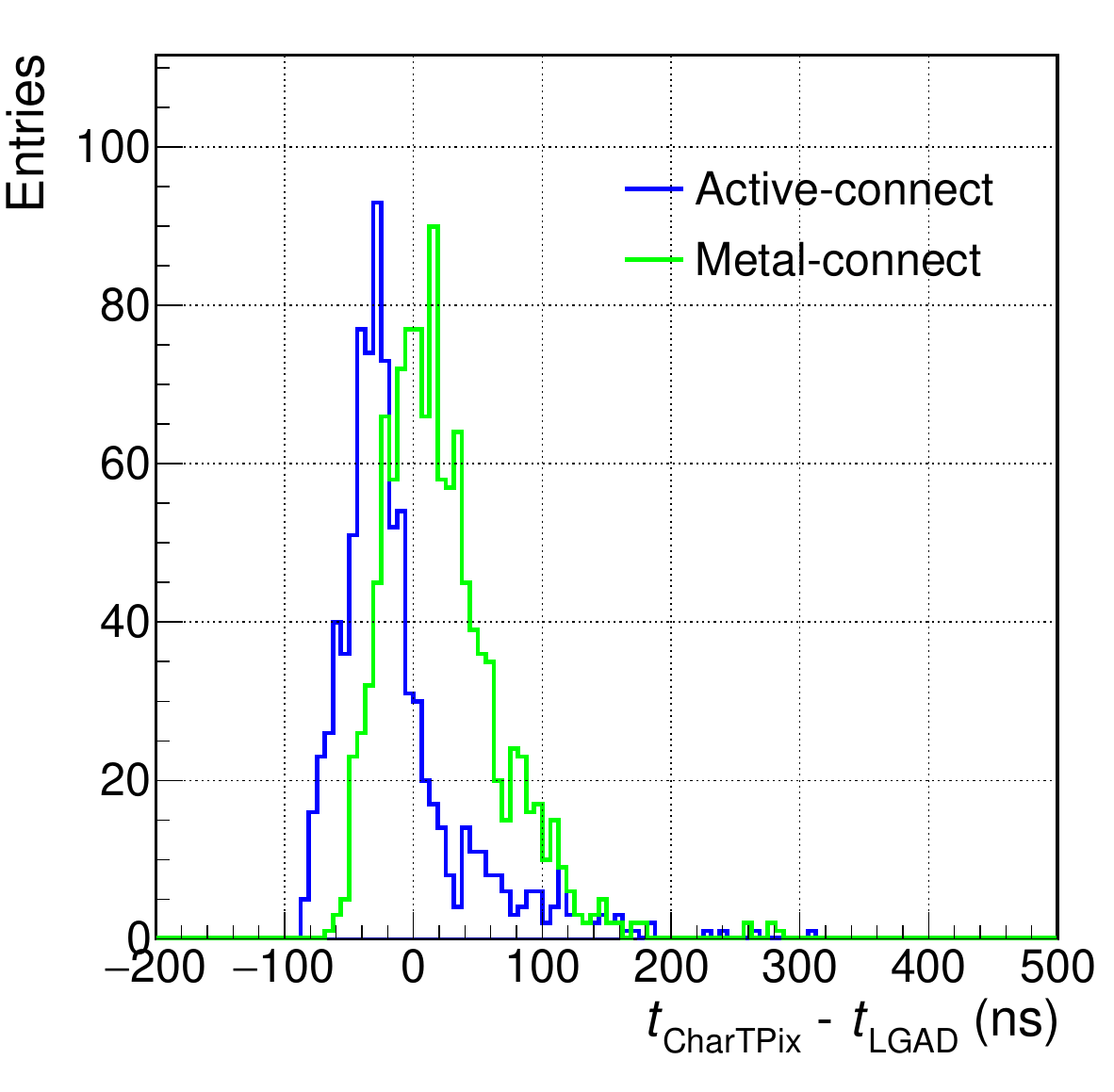}
\caption{Time residual distribution for two sensor types at a typical threshold of \SI{350}{e^-}, measured using a $^{90}$Sr beta source. The substrate bias voltage was set to \SI{-6}{V}.}\label{residual_Sr90}
\end{figure}

\begin{figure}[htbp]
\centering
\includegraphics[width=0.5\textwidth]{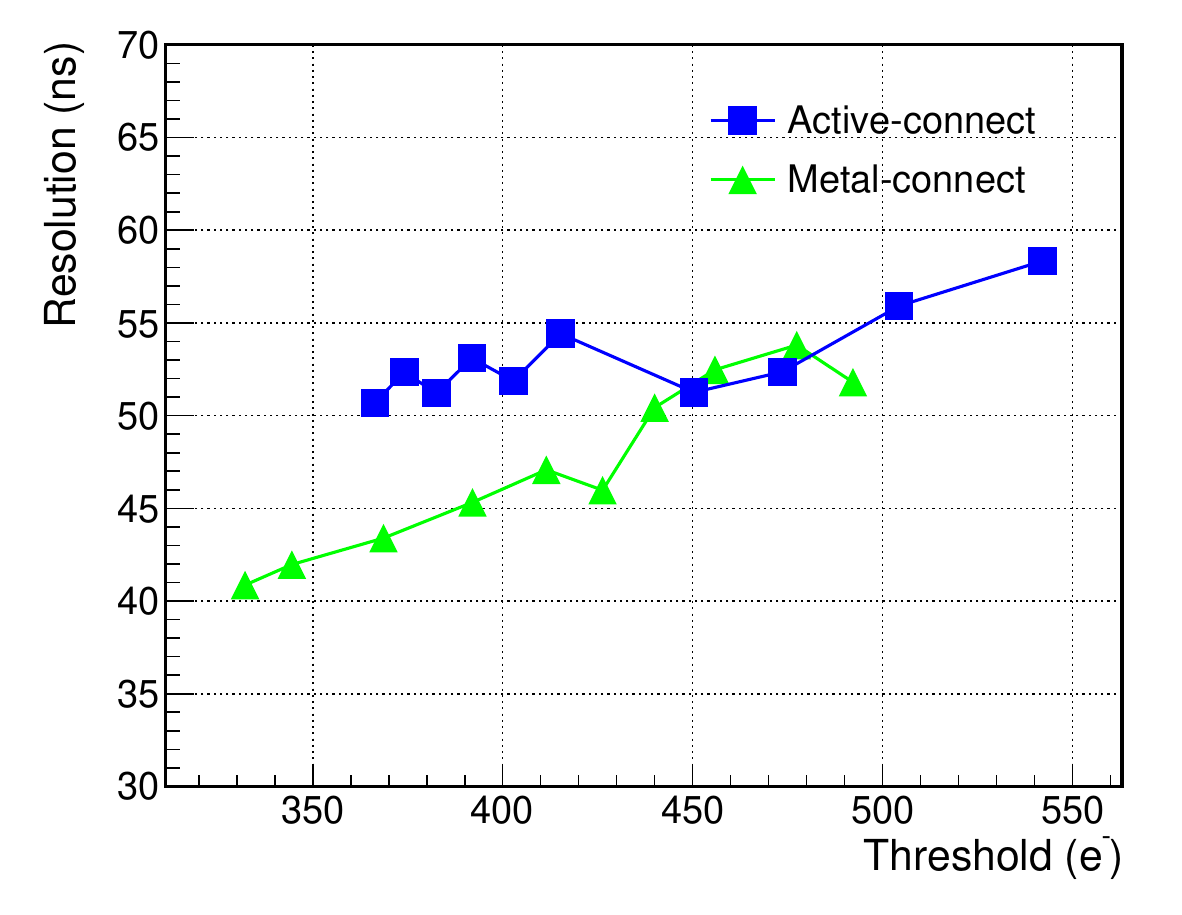}
\caption{Time resolution as a function of the average matrix threshold for two sensor types, measured using a $^{90}$Sr beta source. The substrate bias voltage was set to \SI{-6}{V}.}\label{resolution_Sr90}
\end{figure}

The time resolution of the two sensor types was obtained from the RMS value of their respective time-residual distributions. As shown in Fig.~\ref{residual_Sr90}, the time-residual distribution exhibited a non-Gaussian shape with a long tail, mainly due to the absence of TA calibration. The resulting time resolution as a function of the threshold is presented in Fig.~\ref{resolution_Sr90}.


For both sensor types, the time resolution improved as the threshold was lowered. This threshold dependence should be interpreted as a combined effect of threshold-crossing jitter, time walk, and other contributions, since the measurements were performed without TA calibration. The reduction in threshold-crossing jitter provides a plausible explanation for the observed improvement in time resolution. For a given signal waveform, the jitter induced by voltage fluctuations can be approximately expressed as $\sigma_t \approx \sigma_V/|dV/dt|$, where $\sigma_V$ is the voltage noise and $dV/dt$ is the signal slew rate at the threshold-crossing point. A lower threshold corresponds to an earlier point on the rising edge, where the signal exhibits a larger slew rate and hence a smaller noise-induced timing jitter. 

The best time resolution achieved was \SI{41}{ns} for the metal-connect sensor and \SI{51}{ns} for the active-connect sensor. Interestingly, the active-connect sensor exhibits worse timing performance than the metal-connect sensor in the full-functional chip, in contrast to the results obtained with the analog readout chip presented in Section~\ref{timing_performance}, where the active-connect sensor showed significantly better timing performance. This behavior suggests that the inferior timing performance of the active-connect sensor in the full-functional chip is unlikely to originate primarily from the sensor itself, but rather from an increased contribution from the readout circuitry. This contribution is likely related to the larger sensor capacitance of the active-connect sensor type, which degrade the timing performance of the front-end circuit.

\section{Beam test of CharTPix-180A-c3}
\label{testbeam}
A first beam test campaign has been performed at CERN PS beamline in October, 2025 to study the detailed characteristics of the CharTPix-180A-c3. The experimental setup, analysis method and the results will be discussed in this section.

\subsection{Experimental setup}

The beam test was performed using a nine-layer telescope system, comprising three upstream and three downstream tracking reference layers, one timing reference layer and two device-under-test (DUT) layers. All six tracking reference layers used the JadePix-3 CMOS sensor, a rolling-shutter MAPS prototype developed for the CEPC, with pixel pitches of $\SI{26}{\micro\meter}\times\SI{16}{\micro\meter}$ and $\SI{23.11}{\micro\meter}\times\SI{16}{\micro\meter}$, depending on the sensor sector~\cite{jadepix3, Jadepix3_telescope}.

An LGAD sensor, introduced in Section~\ref{time_res_c3}, was installed downstream to provide an external time reference. The time bin size was set to \SI{4.5}{ns}, corresponding to a $T_0$ resolution of approximately \SI{1.3}{ns}. 

\begin{figure}[htbp]
\centering
\includegraphics[width=1.0\textwidth]{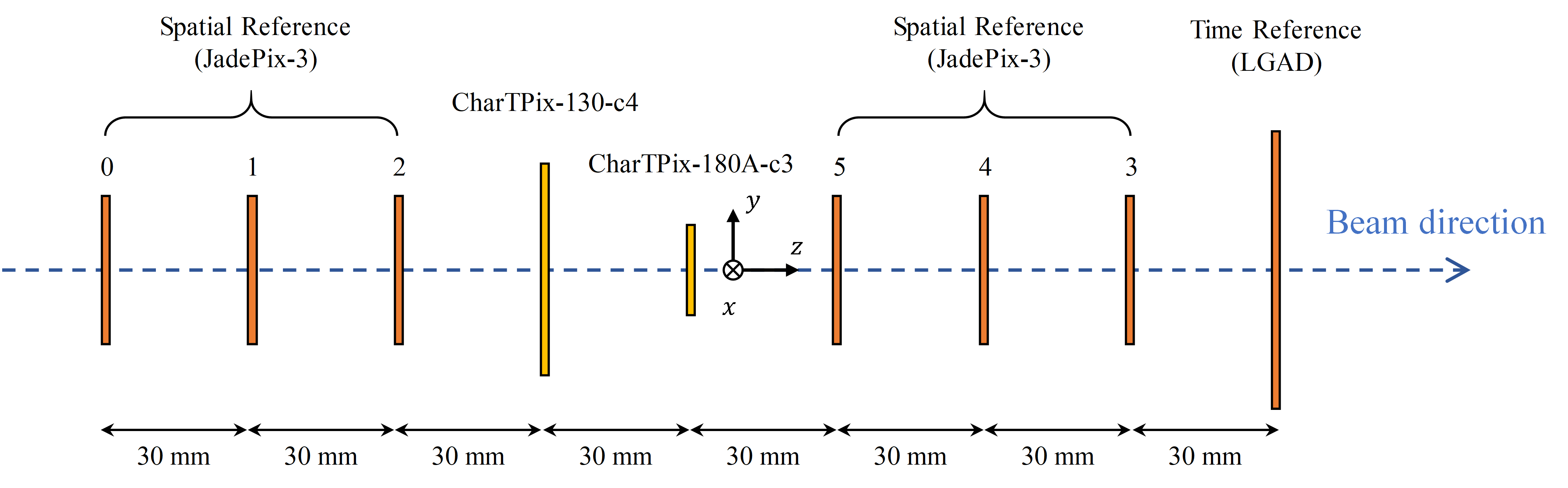}
\caption{Schematic layout of the beam telescope (not to scale). The six JadePix-3 sensors are numbered from 0 to 5 as indicated in the figure.}\label{sketch_telescope}
\end{figure}

The two DUTs were positioned at the center of the telescope. One was the CharTPix-180A-c3 prototype investigated in this work. The other was the CharTPix-130-c4, fabricated in a \SI{130}{nm} CMOS imaging process. The CharTPix-130-c4 adopted a super-pixel architecture targeting low power consumption (<\SI{100}{mW/cm^2}), improved timing performance ($\sim$\SI{20}{ns}), and position resolution ($\sim$\SI{10}{\micro\meter})~\cite{GSMC_super_pixel}. In this study, CharTPix-180A-c3 was treated as the primary DUT, hereafter referred to as the DUT, while CharTPix-130-c4 provided auxiliary spatial and timing information in combination with the JadePix-3 and LGAD reference layers. A schematic layout of the telescope is shown in Fig.~\ref{sketch_telescope}. All detector planes were aligned parallel to each other with an inter-plane spacing of \SI{30}{\milli\meter}.

A common clock generator distributed the system clock to all planes. A daisy-chain link propagated a synchronous start signal from the first plane to the subsequent ones, ensuring inter-plane time synchronization. The fixed propagation delay introduced by the daisy-chain was below \SI{10}{ns} and did not affect the time resolution measurement. Slow control and data acquisition were implemented via the IPbus framework.

\begin{figure}[htbp]
\centering
\includegraphics[width=0.5\textwidth]{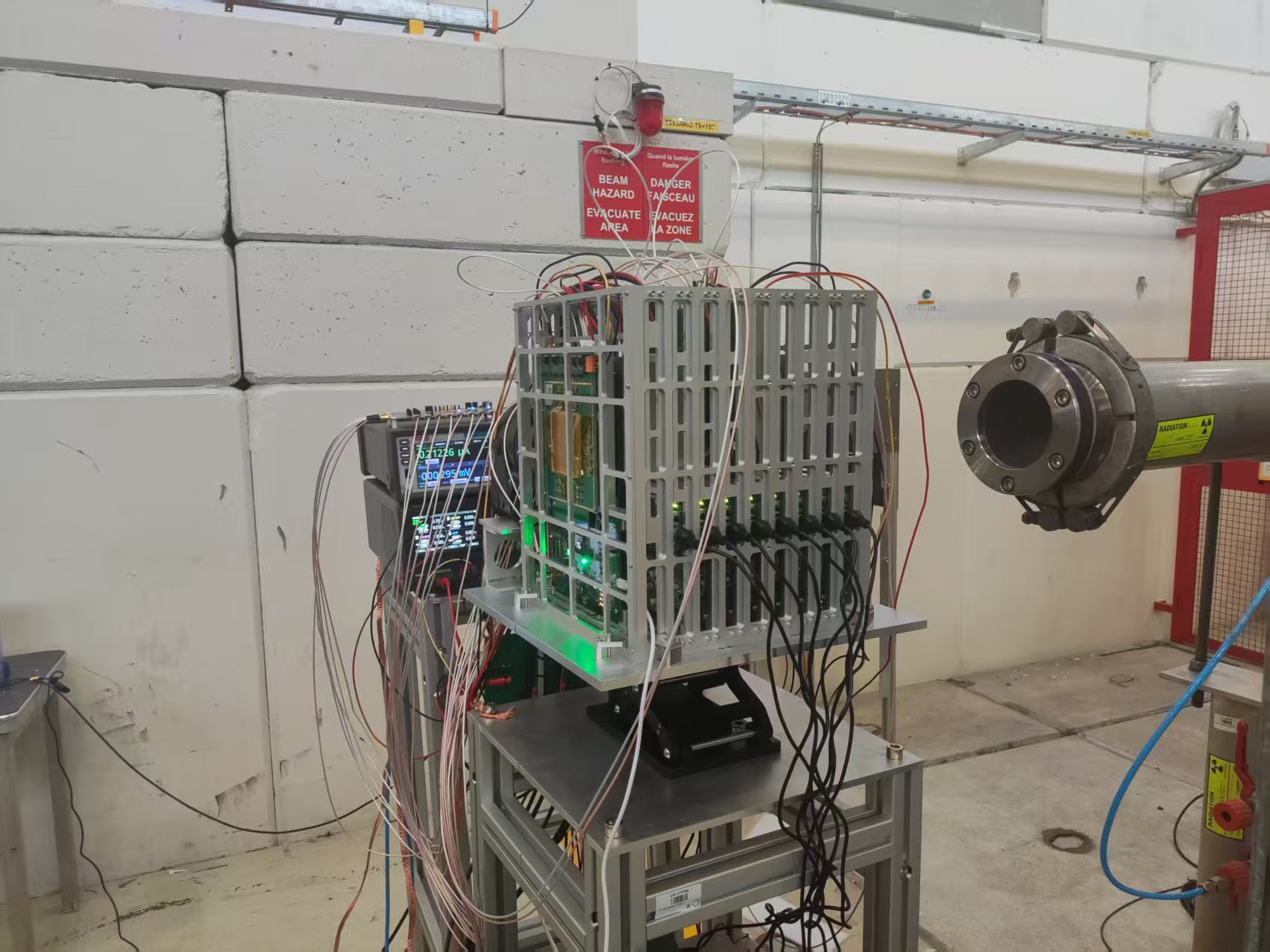}
\caption{Photograph of the beam test setup at the CERN PS T9 beam line. The beam telescope was placed immediately downstream of the beam pipe.}\label{bt_setup}
\end{figure}

A photograph of the experimental setup at the beamline is shown in Fig.~\ref{bt_setup}. The telescope operated at room temperature with air cooling provided by two fans. The beam was directed perpendicularly onto the detector planes from the JadePix-3\_0 side. Due to shared beam time with other experimental groups, hadron beams with momenta of \SI{10}{GeV/c} and \SI{4}{GeV/c} were used during the test. The DUT was characterized under two different substrate bias voltages, \SI{-6}{V} and \SI{-4}{V}. For each bias setting, the beam momentum was kept constant: data taken at \SI{-6}{V} corresponded to a \SI{10}{GeV/c} beam, while measurements at \SI{-4}{V} were performed with a \SI{4}{GeV/c} beam. Various threshold settings were scanned for each operating condition.

\subsection{Analysis method}

The reconstruction and analysis of the test beam data were performed within the Corryvreckan~\cite{corry} framework. Dedicated modules were developed for all four detector types (JadePix-3, CharTPix-180A-c3, CharTPix-130-c4, and LGAD) to enable data decoding and processing.

Events were built using a basic time window of \SI{98.316}{\micro s}, corresponding to the frame duration of the JadePix-3 sensor. Owing to its rolling-shutter readout scheme, hits in JadePix-3 could be assigned to either of two consecutive frame timestamps depending on the row position, leading to a potential time offset relative to the other detector layers. To avoid efficiency loss during event building, the event time window for the three non-JadePix-3 detectors was extended backward by an additional \SI{160}{\micro s}. All hits within the defined time window were considered as possibly originating from the same particle and were subsequently passed to the clustering and track reconstruction procedures.

Clusters in the JadePix-3 and CharTPix-130-c4 detectors were reconstructed using arithmetic and charge-weighted means, respectively. For CharTPix-180A-c3, each cluster consisted of a single pixel hit due to the pixel matrix reset operation; therefore, the hit position was directly used as the cluster position. Track reconstruction was performed using the General Broken Lines (GBL) model~\cite{GBL}. A valid track candidate was required to have associated clusters on all six JadePix-3 reference layers and to satisfy $\chi^2/ndof < 5$. For cluster association on the DUT, a spatial acceptance window of \SI{500}{\micro\meter} around the projected track position was applied.

The detection efficiency of CharTPix-180A-c3 was defined as the fraction of selected tracks with an associated DUT cluster. Tracks passing within one pixel pitch of the masked pixels or traversing the outermost pixel rows and columns were excluded from the analysis. Furthermore, CharTPix-180A-c3 exhibited a chip-level dead time (cf. Section~\ref{chip3}), during which a track occurring within a certain time interval after a preceding hit could not produce a valid response, resulting in an artificial loss of detection efficiency. Such tracks were identified and excluded from the efficiency calculation using track timestamp information. The track timestamp was derived from the CharTPix-130-c4 layer, which provided both good spatial resolution ($\sim$\SI{10}{\micro\meter}) and adequate time resolution ($\sim$\SI{50}{ns}).\footnote{The LGAD was not used for track timestamping due to its limited spatial granularity, which prevented unambiguous timing assignment when multiple tracks were present within the same event---a frequent occurrence under the high-rate beam conditions.} Consequently, only tracks with an associated CharTPix-130-c4 cluster were considered in the efficiency analysis. A dead time interval of \SI{30}{\micro s} following each valid track was applied when rejecting subsequent tracks.

For the spatial resolution evaluation, CharTPix-180A-c3 clusters located within 1.5 pixel pitches of the projected track position were considered. The residual distributions were fitted with Gaussian functions, and the standard deviation was taken as the measured spatial resolution. The extrapolated track uncertainty, estimated using the GBL model, was then subtracted in quadrature to obtain the intrinsic position resolution of the DUT. For a beam momentum of \SI{10}{GeV/c}, the extrapolated track uncertainties were 
$\sigma_{\text{track,col}} = \SI{5.6}{\micro\meter}$ and 
$\sigma_{\text{track,row}} = \SI{4.5}{\micro\meter}$. 
For \SI{4}{GeV/c}, the corresponding values were 
$\sigma_{\text{track,col}} = \SI{8.6}{\micro\meter}$ and 
$\sigma_{\text{track,row}} = \SI{7.6}{\micro\meter}$.

The time resolution was evaluated using the LGAD as a reference. Time residuals were calculated within a window of \SI{-350}{ns} to \SI{100}{ns}, optimized to suppress accidental coincidences while retaining true correlated hits. Owing to the asymmetric shape of the residual distribution, the time resolution was determined from the standard deviation without applying a Gaussian fit.


A fake-hit rate measurement was performed using the same dataset acquired during beam operation. The beam exhibited a periodic structure consisting of a $\sim$\SI{1}{s} spill period followed by a $\sim$\SI{30}{s} idle period. DUT hits recorded during the idle period were used to evaluate the fake-hit rate. The boundary between spill and idle periods was determined from the nearly noise-free LGAD time distribution. To avoid contamination from beam particles due to timing misalignment, an additional \SI{100}{ms} margin was excluded on both sides of the boundary. To suppress contributions from hot pixels, pixels with an idle-period hit rate exceeding \SI{10}{Hz} were identified as hot. These pixels were ranked in descending order of their hit rate and sequentially masked. For each pixel variant, the total fraction of masked pixels was limited to less than 1\% of the corresponding number of pixels.

\subsection{Result and discussion}

\begin{figure}[htbp]
\centering
\includegraphics[width=1.0\textwidth]{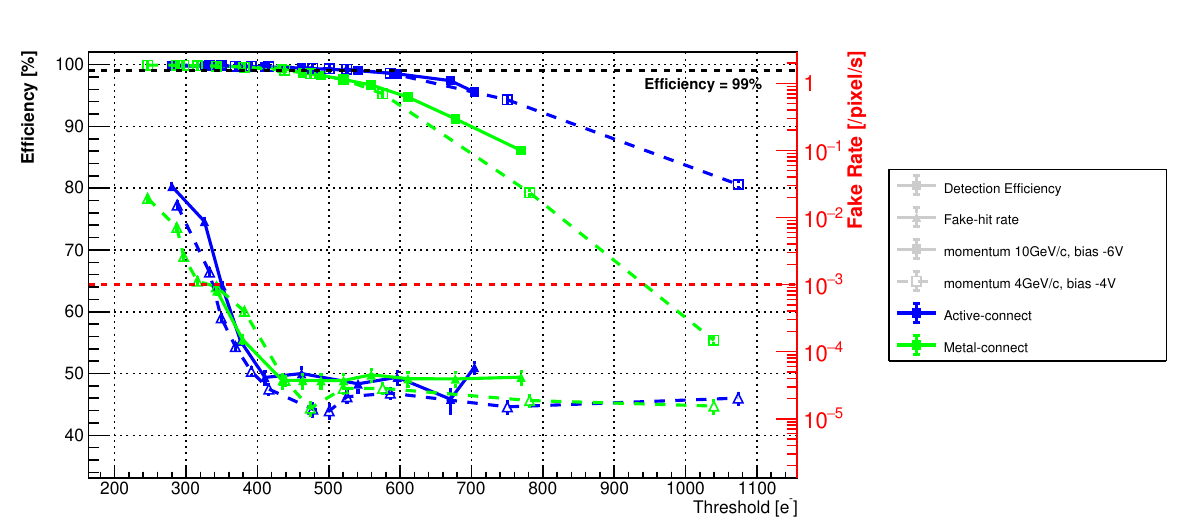}
\caption{Detection efficiency and fake-hit rate as functions of the average matrix threshold for the active-connect and metal-connect sensor types under different beam and bias conditions.}\label{eff_fakerate}
\end{figure}

Fig.~\ref{eff_fakerate} shows the detection efficiency and fake-hit rate as functions of the average matrix threshold (measured in situ) for the two sensor types. According to the STCF inner tracker requirements, the MAPS device is required to operate with a detection efficiency above 99\% while maintaining a fake-hit rate below \SI{1e-3}{/pixel/s}. These performance criteria are indicated by two dashed lines in the figure.

The detection efficiency remained above 99\% up to a threshold of approximately \SI{500}{e^-} for both sensor types. Beyond this value, the efficiency decreased, with a more pronounced degradation observed for the metal-connect sensor at higher thresholds. This behavior was consistent with the laboratory measurements using $^{55}$Fe and with previous simulation studies. When comparing data taken at different beam momenta and substrate bias voltages, the active-connect sensor exhibited negligible variation. In contrast, the metal-connect sensor showed improved efficiency under the condition of \SI{10}{GeV/c} beam momentum and a substrate bias of \SI{-6}{V}. Since particles at both beam momenta can be regarded as minimum ionizing particles (MIPs) and are not expected to significantly affect charge deposition, the observed difference was attributed primarily to the variation in substrate bias. The results indicate that increasing the bias voltage from \SI{-4}{V} to \SI{-6}{V} enhanced the performance of the metal-connect design, whereas the active-connect design was largely insensitive to this bias change within the studied range. This behavior can be explained by the extended p--n junction structure of the active-connect variant, which enables full depletion at relatively low substrate bias.

The fake-hit rate decreased exponentially with increasing threshold and reached a stable level of approximately \SI{2e-5}{/pixel/s} at around \SI{400}{e^-} for both sensor types. No significant dependence on substrate bias was observed. The achieved fake-hit rate was of the same order of magnitude as that reported for ALPIDE~\cite{alpide}. Considering both efficiency and fake-hit constraints, an operational margin of roughly \SI{200}{e^-} was obtained for both sensor types.

\begin{figure}[htbp]
\centering
    \begin{subfigure}{0.6\textwidth}
        \centering
        \includegraphics[width=\linewidth]{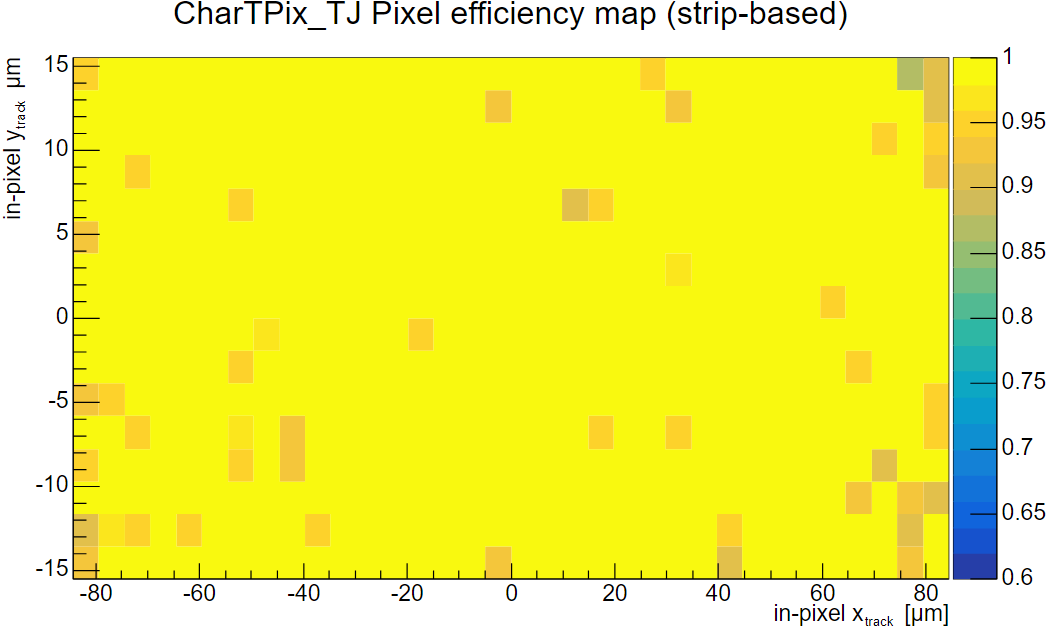}
        \caption{active-connect}
        \label{eff_active}
    \end{subfigure}
    \hfill
    \begin{subfigure}{0.6\textwidth}
        \centering
        \includegraphics[width=\linewidth]{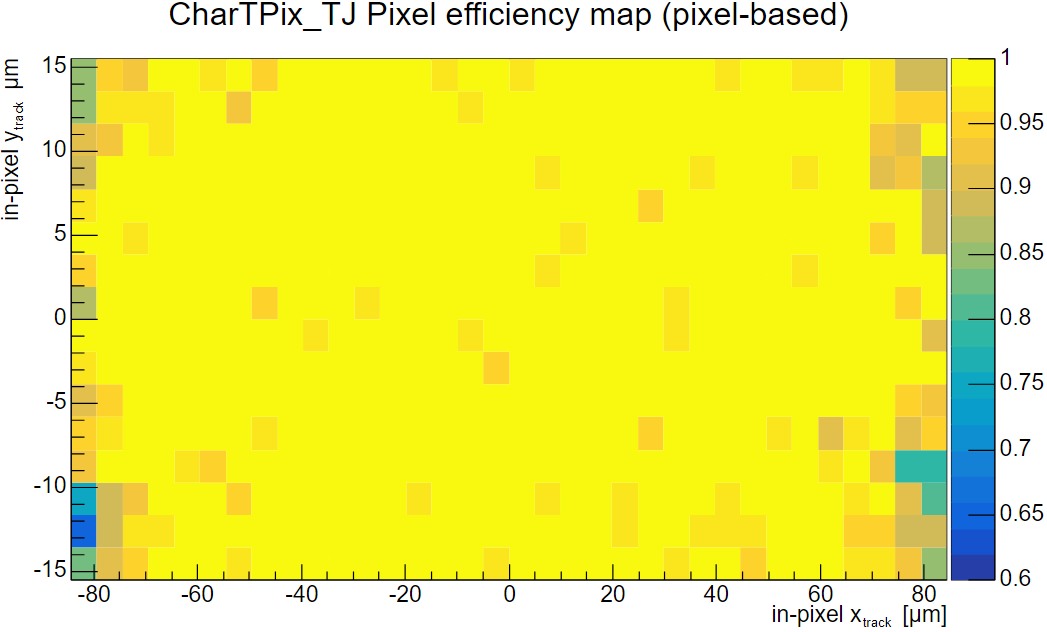}
        \caption{metal-connect}
        \label{eff_metal}
    \end{subfigure}
\caption{In-pixel detection efficiency maps for (a) the active-connect sensor and (b) the metal-connect sensor, measured with a \SI{10}{GeV/c} hadron beam at a substrate bias of \SI{-6}{V}. The threshold was set to \SI{460}{e^-}.}\label{eff_in_pixel}
\end{figure}

The in-pixel detection efficiency distributions are presented in Fig.~\ref{eff_in_pixel} for a representative operating point of \SI{460}{e^-} threshold and \SI{-6}{V} substrate bias. Although the statistical precision was limited by the small chip size and available beam time, a reduction in efficiency near the pixel corners was observable. This effect was more pronounced for the metal-connect pixel type, consistent with expectations based on its electrode geometry.

\begin{figure}[htbp]
\centering
    \begin{subfigure}{0.6\textwidth}
        \centering
        \includegraphics[width=\linewidth]{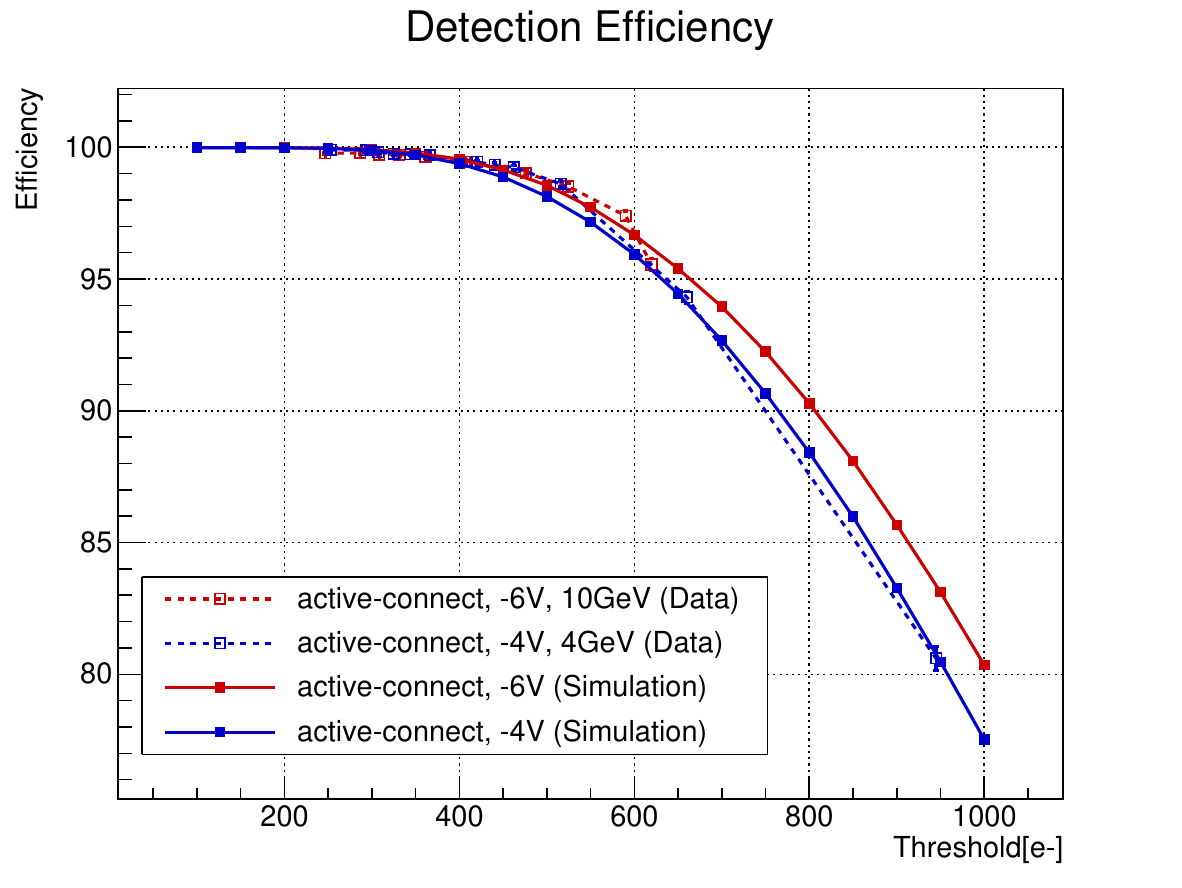}
        \caption{active-connect, scale factor = 0.88}
    \end{subfigure}
    \hfill
    \begin{subfigure}{0.6\textwidth}
        \centering
        \includegraphics[width=\linewidth]{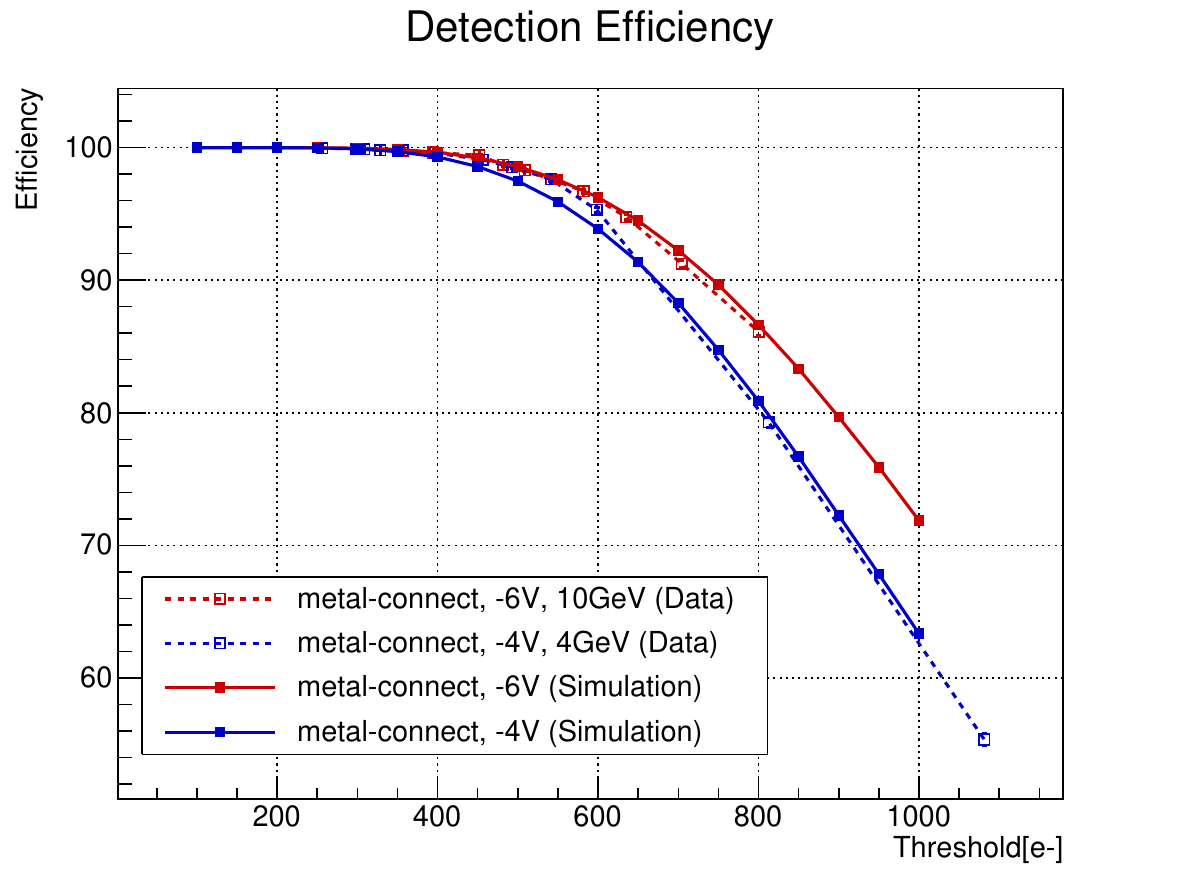}
        \caption{metal-connect, scale factor = 1.04}
    \end{subfigure}
\caption{Comparison between measured and simulated detection efficiency as functions of threshold for (a) the active-connect sensor and (b) the metal-connect sensor. The experimental threshold values were rescaled using constant calibration factors (see text for details).}\label{eff_sim}
\end{figure}

A comparison between the measured efficiency curves and simulation was subsequently performed. The simulation procedure followed the MIP modeling approach described in~\cite{stcf_maps_sim}, with the sensor structure updated to match that used in Section~\ref{c4_fe55}. An increasing discrepancy between simulation and measurement was observed at higher thresholds. This deviation was attributed to a systematic bias in the threshold calibration, originating from uncertainties in the injection capacitance used for charge conversion. To account for this effect, the experimental threshold values were rescaled by applying a constant multiplicative factor to achieve optimal agreement between simulation and data. Since different injection capacitances were employed for the two sensor matrices, the scaling factors were determined separately. The final factors were 0.88 for the active-connect sensor and 1.04 for the metal-connect sensor. The comparison after calibration is shown in Fig.~\ref{eff_sim}. Good agreement between simulation and measurement was achieved over the full threshold range, supporting the validity of the adopted simulation model.

\begin{figure}[htbp]
\centering
    \begin{subfigure}{0.5\textwidth}
        \centering
        \includegraphics[width=\linewidth]{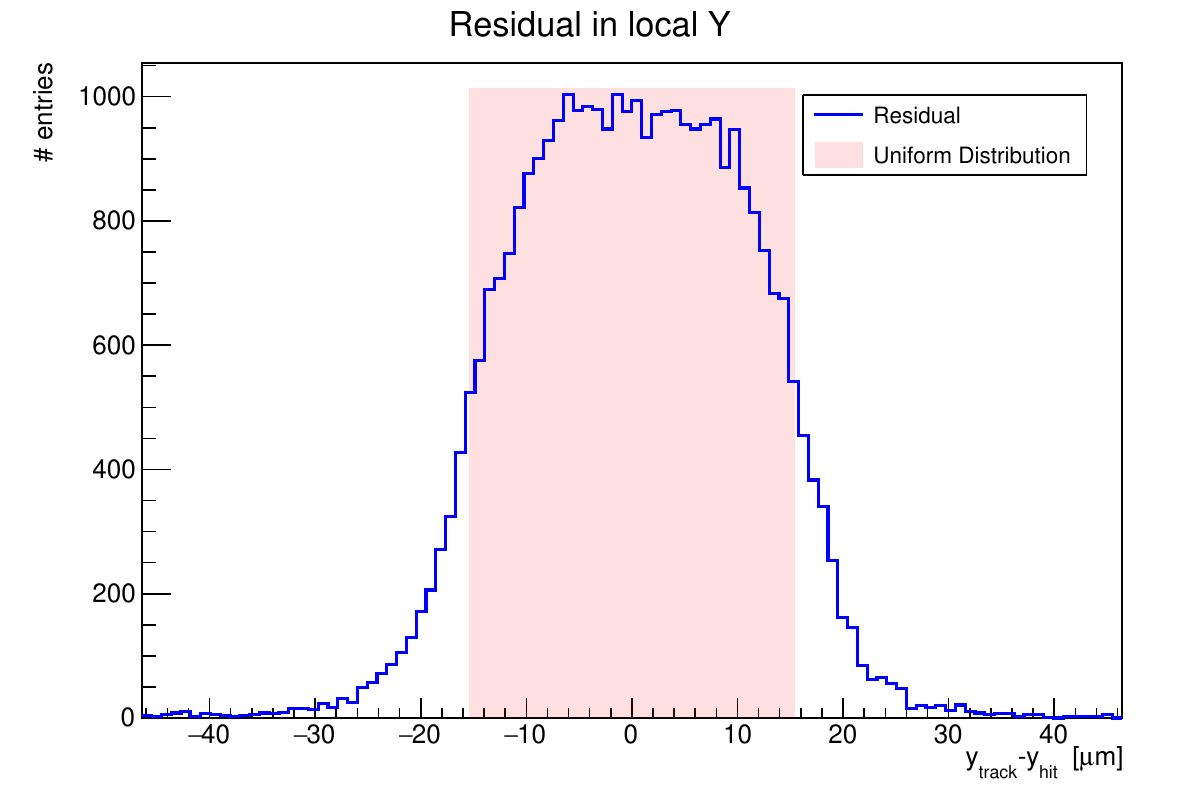}
        \caption{Row direction, pixel pitch = \SI{31}{\micro\meter}}
        \label{spatial_residual_row}
    \end{subfigure}
    \hfill
    \begin{subfigure}{0.5\textwidth}
        \centering
        \includegraphics[width=\linewidth]{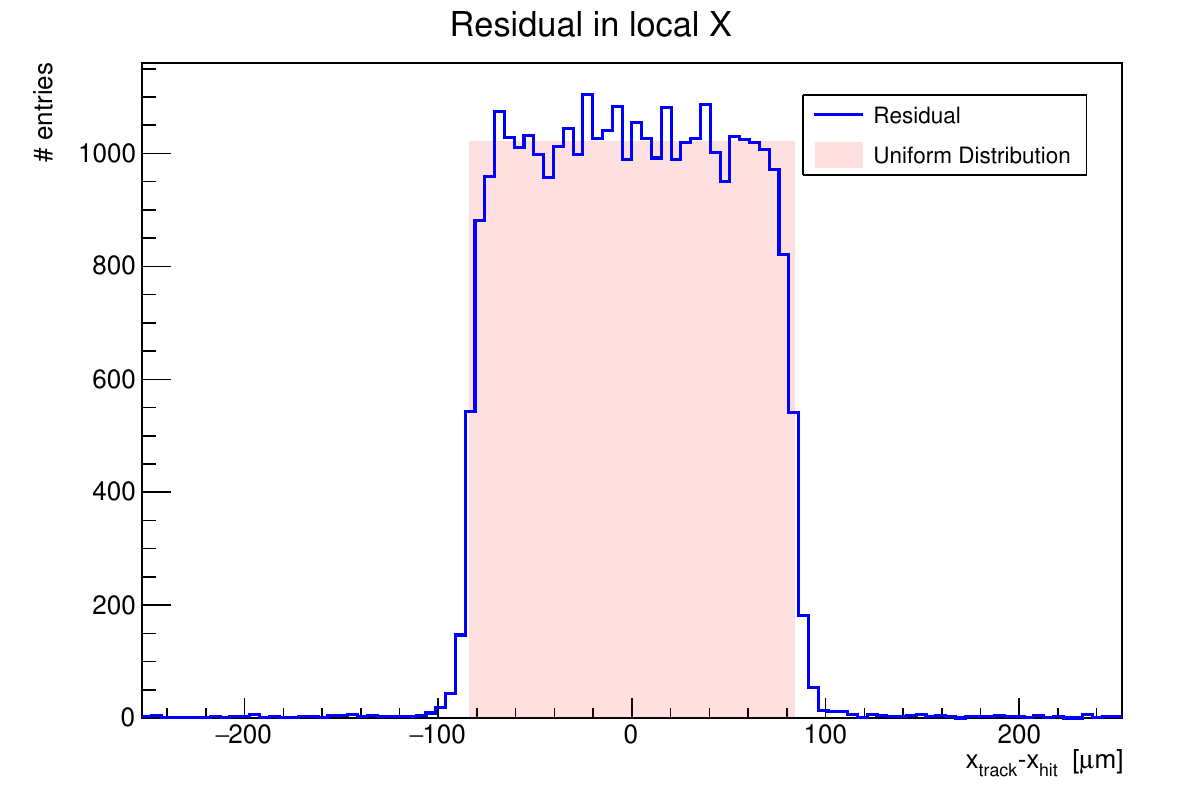}
        \caption{Column direction, pixel pitch = \SI{170}{\micro\meter}}
        \label{spatial_residual_column}
    \end{subfigure}
\caption{The spatial residual distribution along (a) the row direction and (b) the column direction for active-connect sensor at a threshold of \SI{460}{e^-}, measured with \SI{10}{GeV/c} hadron beam at \SI{-6}{V} bias voltage.}\label{spatial_residual}
\end{figure}

\begin{figure}[htbp]
\centering
    \begin{subfigure}{0.6\textwidth}
        \centering
        \includegraphics[width=\linewidth]{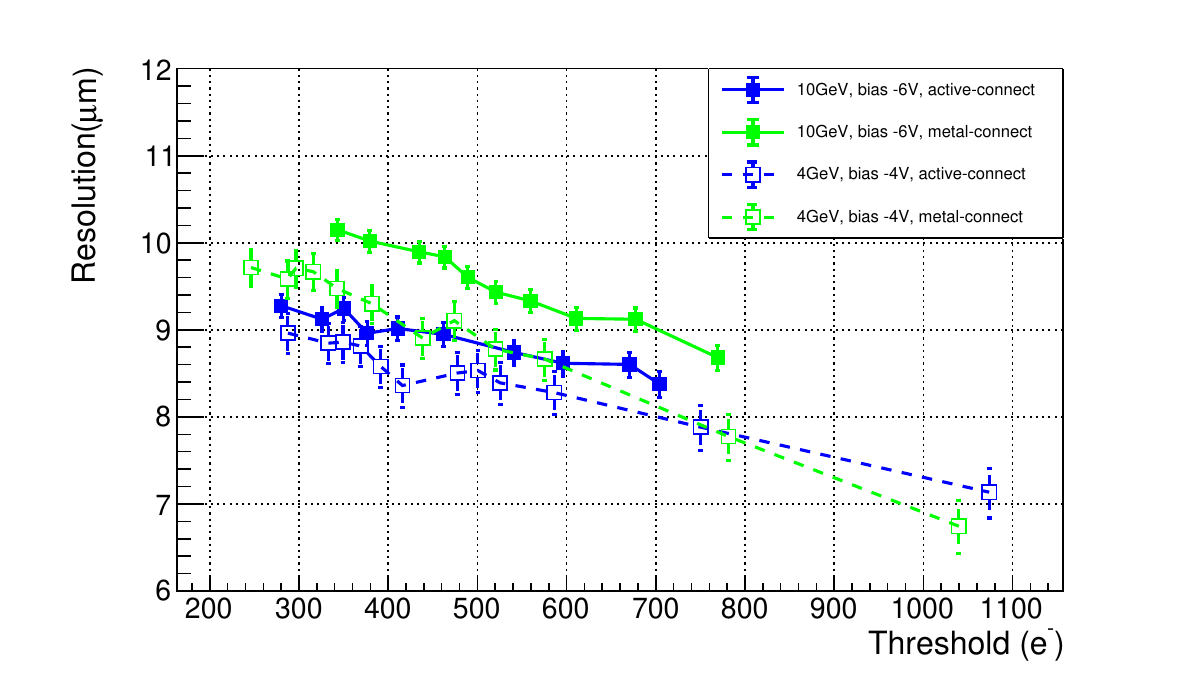}
        \caption{Row direction, pixel pitch = \SI{31}{\micro\meter}}
        \label{spatial_resolution_row}
    \end{subfigure}
    \hfill
    \begin{subfigure}{0.6\textwidth}
        \centering
        \includegraphics[width=\linewidth]{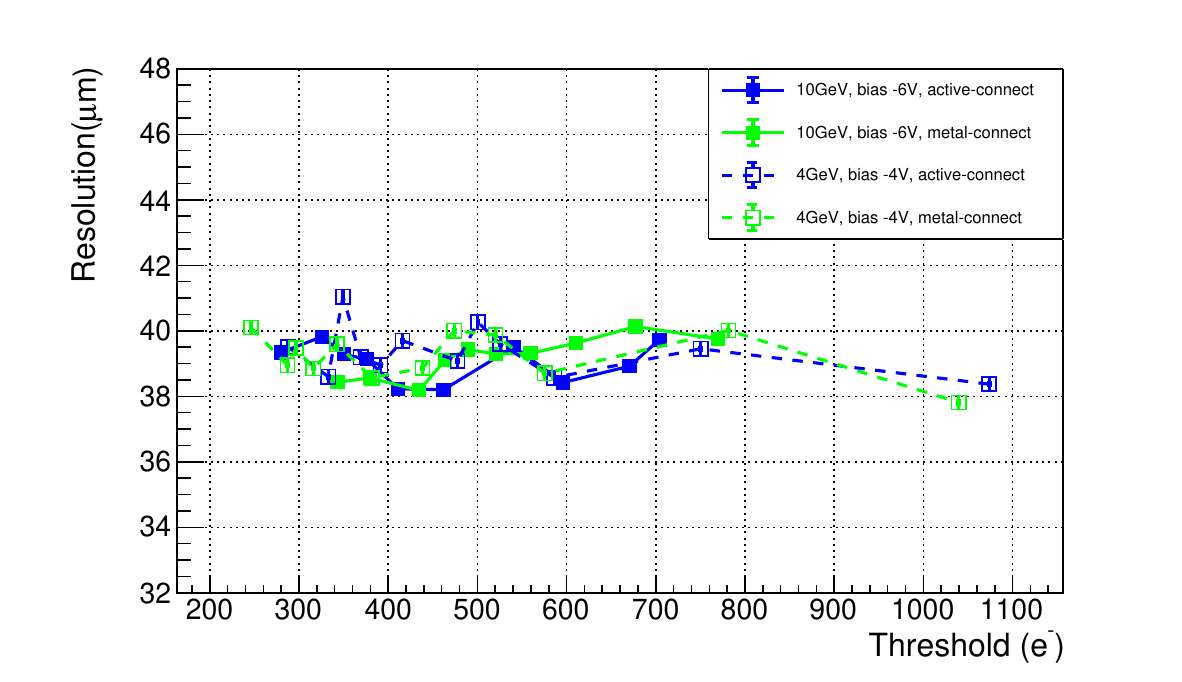}
        \caption{Column direction, pixel pitch = \SI{170}{\micro\meter}}
        \label{spatial_resolution_col}
    \end{subfigure}
\caption{The spatial resolution along (a) the row direction and (b) the column direction for the two sensor types under different experimental conditions.}\label{spatial_resolution}
\end{figure}

The spatial residual distributions for the active-connect sensor at a substrate bias of \SI{-6}{V} and a representative threshold of \SI{460}{e^-} are displayed in Fig.~\ref{spatial_residual}, with the row and column directions shown separately. For comparison, a uniform distribution spanning the corresponding pixel pitch was superimposed on each residual distribution. The calculated spatial resolution as a function of the average threshold under different experimental conditions is presented in Fig.~\ref{spatial_resolution}.

In the row direction, corresponding to a fine pixel pitch of \SI{31}{\micro\meter}, the spatial resolution was measured to be better than \SI{10}{\micro\meter} and showed a gradual improvement with increasing threshold. This behavior originated from the limitation that only a single pixel within each cluster could be read out. At low threshold, the intrinsic cluster size in the row direction was typically larger than one pixel due to charge sharing. Since the readout logic randomly selected one of the fired pixels, an additional uncertainty was introduced in the hit position assignment, leading to a degradation of the resolution. As the threshold increased, the effective cluster size approached unity, reducing this ambiguity and allowing the resolution to converge toward the value expected for a properly functioning device. Consequently, the spatial residual distribution deviated from the ideal Gaussian shape anticipated for charge-centroid reconstruction, as illustrated in Fig.~\ref{spatial_residual_row}.

In contrast, along the column direction with a pixel pitch of \SI{170}{\micro\meter}, the measured resolution was approximately \SI{39}{\micro\meter} and exhibited no clear dependence on threshold. This distinct behavior was attributed to the substantially larger pixel pitch in this direction. Charge sharing between neighboring pixels was limited, and the intrinsic cluster size remained close to one even at low threshold. As a result, the impact of the readout constraint on position determination was minimal. The corresponding residual distribution, shown in Fig.~\ref{spatial_residual_column}, is consistent with a nearly uniform shape. Under such conditions, the resolution extracted from a Gaussian fit should be regarded as a qualitative indicator rather than a precise statistical measure, which accounts for the relatively large fluctuations observed in Fig.~\ref{spatial_resolution_col}. Nevertheless, the achieved column direction resolution already satisfied the \SI{100}{\micro\meter} requirement specified for the STCF inner tracker application~\cite{CDR}.

\begin{figure}[htbp]
\centering
\includegraphics[width=1.0\textwidth]{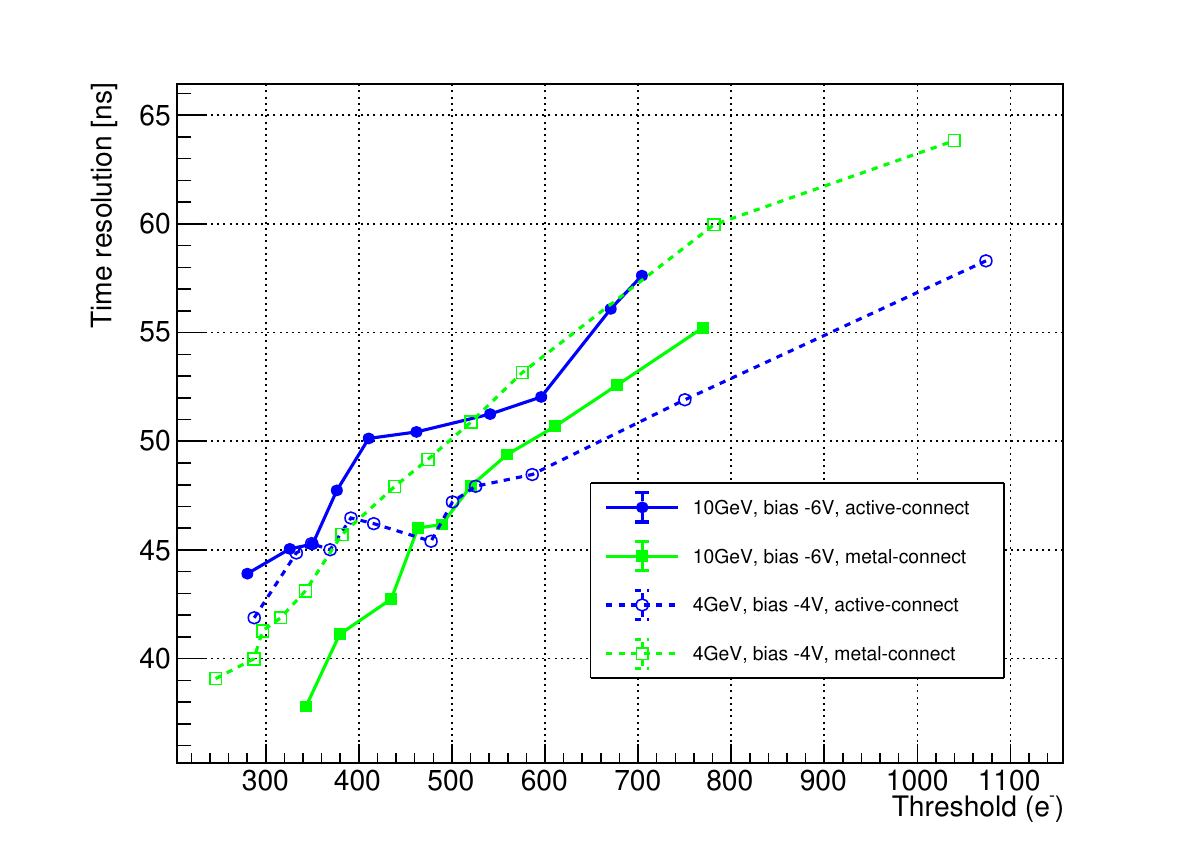}
\caption{Time resolution as a function of the average matrix threshold for the active-connect and metal-connect sensor types under different beam and bias conditions.}\label{time_resolution_tb}
\end{figure}

Fig.~\ref{time_resolution_tb} shows the measured time resolution as a function of the average threshold under different experimental conditions. The resolution varied between \SI{37}{ns} and \SI{64}{ns}, with the best value of \SI{37}{ns} achieved for the metal-connect sensor at a substrate bias of \SI{-6}{V}. A clear trend of improved time resolution at lower operational thresholds was observed for both sensor types. This behavior was consistent with laboratory measurements and was attributed to the more favorable operating conditions of the front-end electronics at reduced thresholds, where signal amplitudes more effectively surpass the discrimination level. When comparing the two substrate bias settings, the metal-connect sensor exhibited better timing performance at \SI{-6}{V}, which is consistent with expectations based on faster charge collection under increased depletion. In contrast, the active-connect sensor showed the opposite trend. The origin of this discrepancy remains unclear and may be related either to variations in beam conditions or to differences in front-end response. Further dedicated characterization will be required to clarify this behavior. It should be noted that TA calibration was not feasible in this prototype. Therefore, the measured values do not yet represent the intrinsic limit of the sensor and front-end design. Improved timing performance is anticipated in the next chip iteration, with the aim of achieving the target resolution of \SI{20}{ns}.

\section{Conclusion and outlook}

Two prototype chips of the CharTPix series fabricated in a \SI{180}{nm} CMOS imaging process were systematically characterized. Laboratory measurements of the analogue readout prototype, CharTPix-180A-c4, enabled a detailed comparison between the active-connect and metal-connect sensor variants, including sensor capacitance and charge collection performance. These studies provided essential insight into the intrinsic characteristics of the two sensor variants and offered quantitative guidance for subsequent design optimization. The digital readout prototype, CharTPix-180A-c3, was evaluated through both laboratory and beam tests. Its fundamental functionality was successfully demonstrated, and a comprehensive performance assessment was carried out. The estimated power densities were \SI{54.7}{mW/cm^2} for the active-connect variant and \SI{42.1}{mW/cm^2} for the metal-connect variant, with the latter meeting the STCF target of \SI{50}{mW/cm^2} before accounting for additional readout and peripheral power contributions. A detection efficiency exceeding 99\% with a fake-hit rate below \SI{1e-3}{/pixel/s} was achieved for both sensor types, with an operational threshold margin of approximately \SI{200}{e^-}. The measured spatial resolution satisfied the requirements of the STCF inner tracker in both directions. However, the current timing performance remains inferior to that anticipated from the analog readout chip measurements. This limitation is partially attributed to the absence of time-walk calibration, which should be addressed in future chip iterations.

The characterization campaign also served to validate and refine the previously developed simulation framework. After tuning of key model parameters based on measurement results, good agreement between simulation and experiment was achieved in both the X-ray spectrum and the MIP detection efficiency, thereby strengthening confidence in the predictive capability of the model.

Future studies will first focus on evaluating the radiation tolerance of the current prototypes. Systematic irradiation studies under various total ionizing dose (TID) and non-ionizing energy loss (NIEL) conditions will be performed to quantify performance degradation and establish the radiation tolerance of the sensors under conditions relevant to the STCF environment.

In parallel, building on the experience gained from this first prototype generation, an improved version of CharTPix-180A-c3 has been designed and submitted for fabrication using the same process. Previously identified issues, including crosstalk and encoding errors, have been addressed in the updated design. The next iteration of prototype is expected to provide improved overall performance and offer further validation of the design approach.


\section*{Declaration of generative AI and AI-assisted technologies in the writing process}

During the preparation of this work the authors used ChatGPT in order to improve the readability and language of the manuscript. After using this tool/service, the authors reviewed and edited the content as needed and take full responsibility for the content of the published article.

\section*{Acknowledgements}
This work was supported by the National Key Research and Development Program of China under Grant 2022YFA1602203, the National Natural Science Foundation of China under Grant 12341502 and 12125505, and the STCF Key Technology Research and Development Project. We thank the Hefei Comprehensive National Science Center for its strong support on the promotion and development of the STCF project. We would like to express our special thanks to Yunpeng Lu and Sheng Dong for their strong support and valuable guidance on the JadePix-3 chip and telescope system. We also thank Yuedong Sun for his assistance with the design and fabrication of the telescope system, and Yunlong Zhang and Dipanwita Banerjee for their help during the beam test.




\bibliographystyle{elsarticle-num} 
\bibliography{biblio.bib}







\end{document}